\documentclass[pra, aps,letterpaper, preprintnumbers,superscriptaddress]{revtex4}
\usepackage{hyperref}
\usepackage{verbatim}
\usepackage{amsmath}
\usepackage{latexsym}
\usepackage{revsymb}
\usepackage{yfonts}
\usepackage{ifthen}

\usepackage{amsfonts}
\usepackage{amsmath}
\usepackage{amssymb}
\usepackage{amsthm}
\usepackage{graphicx}
\usepackage{bm}
\usepackage{bbm}

\newcommand{\be}{\begin{eqnarray} \begin{aligned}}
\newcommand{\ee}{\end{aligned} \end{eqnarray} }
\newcommand{\benn}{\begin{eqnarray*} \begin{aligned}}
\newcommand{\eenn}{\end{aligned} \end{eqnarray*} }

\newcommand{\Mat}{\mathbb{M}}
\newcommand{\Enc}{\mathrm{Enc}}
\newcommand{\Syn}{\mathrm{Syn}}
\newcommand{\Ext}{\mbox{Ext}}
\newcommand{\isomorph}{\simeq}

\newcommand{\bc}{\begin{center}}
\newcommand{\ec}{\end{center}}

\newcommand{\id}{\mathbb{I}}

\newcommand{\Tr}{\mathop{\mathrm{Tr}}\nolimits}

\newcommand{\e}{\mathrm{e}}

\newtheorem{theorem}{Theorem}[section]
\newtheorem{claim}[theorem]{Claim}

\newtheorem{lemma}[theorem]{Lemma}
\newtheorem{definition}[theorem]{Definition}

\newtheorem{corollary}[theorem]{Corollary}

\newcommand{\bop}{\mathcal{B}}

\newcommand{\hout}{\mathcal{H}_{\rm out}}
\newcommand{\hin}{\mathcal{H}_{\rm in}}

\newcommand*{\cA}{\mathcal{A}} 
\newcommand*{\cB}{\mathcal{B}}

\newcommand*{\cE}{\mathcal{E}}
\newcommand*{\cF}{\mathcal{F}}

\newcommand*{\cH}{\mathcal{H}}
\newcommand*{\cI}{\mathcal{I}}

\newcommand*{\cK}{\mathcal{K}}

\newcommand*{\cM}{\mathcal{M}}
\newcommand*{\cN}{\mathcal{N}}

\newcommand*{\cR}{\mathcal{R}}

\newcommand*{\cS}{\mathcal{S}}

\newcommand*{\cT}{\mathcal{T}}

\usepackage{amsfonts}

\def\id{\mathbb{I}}

\def\01{\{0,1\}}
\newcommand*{\sbin}{\{0,1\}}

\newcommand{\eps}{\varepsilon}
\newcommand{\ket}[1]{|#1\rangle}
\newcommand{\bra}[1]{\langle#1|}
\newcommand{\proj}[1]{|#1\rangle\langle#1|}
\newcommand{\outp}[2]{|#1\rangle\langle#2|}

\newcommand{\assign}{\ensuremath{\kern.5ex\raisebox{.1ex}{\mbox{\rm:}}\kern -.3em =}}

\newcommand{\src}{{\rm src}}
\newcommand{\sent}{{\rm sent}}
\newcommand{\sents}{{{\rm sent},s}}

\newcommand{\sclickA}{{\rm A,S,click}}

\newcommand{\sig}{{\rm sig}}
\newcommand{\vac}{{\rm vac}}
\newcommand{\clickB}{{\rm B,click}}
\newcommand{\clickBs}{{{\rm B,click},s}}
\newcommand{\sclickB}{{\rm B,S,click}}
\newcommand{\dclickB}{{\rm B,D,click}}
\newcommand{\clickA}{{\rm A,click}}
\newcommand{\lossC}{{\rm erase}}

\newcommand{\lostB}{{\rm B,no\ click}}
\newcommand{\lostBs}{{{\rm B,no\ click},s}}
\newcommand{\slostB}{{\rm B,S,no\ click}}
\newcommand{\dlostB}{{\rm B,D,no\ click}}

\newcommand{\errB}{{\rm B, err}}
\newcommand{\serrB}{{\rm B,S, err}}
\newcommand{\snoerrB}{{\rm B,S,no\ err}}

\newcommand{\dark}{{\rm dark}}

\newcommand{\serr}{{\rm B,S,err}}
\newcommand{\derr}{{\rm B,D, err}}
\newcommand{\dserr}{{\rm B,DS,err}}
\newcommand{\edet}{{e_{\rm det}}}

\newcommand{\pdark}{p_{\rm dark}}

\newcommand{\syn}{{\rm syn}}

\renewcommand{\H}{\operatorname{H}} 
\newcommand{\hmin}{\ensuremath{\H_{\infty}}}

\newcounter{protoCount}
\newcounter{protoList}
\newsavebox{\tmpbox}
\newlength{\protobox}
\newenvironment{protocol}[3]{
\bigskip
\addtocounter{protoCount}{1}
\noindent \begin{lrbox}{\tmpbox}
\setlength{\protobox}{\textwidth}
\addtolength{\protobox}{-0.5cm}
\begin{minipage}[c]{\protobox}
\begin{bfseries}Protocol #1: #2\end{bfseries}
\ifthenelse{\equal{#3}{\empty}}{}{\\ #3}
\begin{list}{\begin{bfseries}\arabic{protoList}:\end{bfseries}}
{\usecounter{protoList}}
}{
\end{list}
\end{minipage}\end{lrbox}
\fbox{\usebox{\tmpbox}}
\bigskip
}

\begin{document}

\title{Implementation of two-party protocols in the noisy-storage model}

\author{Stephanie Wehner}
\affiliation{Institute for Quantum Information, Caltech, Pasadena, CA 91125, USA}
\email{wehner@caltech.edu}
\author{Marcos Curty}
\affiliation{ETSI Telecomunicaci\'on, Department of Signal Theory and Communications, University of Vigo, 
Campus Universitario, E-36310 Vigo (Pontevedra), Spain}
\email{mcurty@com.uvigo.es}
\author{Christian Schaffner}
\affiliation{Centrum Wiskunde \& Informatica (CWI), P.O. Box 94079,
  1090 GB Amsterdam, Netherlands}
\email{c.schaffner@cwi.nl}
\author{Hoi-Kwong Lo}
\affiliation{Center for Quantum Information and Quantum Control (CQIQC), 
Department of Physics and Department of Electrical \& Computer Engineering, 
University of Toronto, Toronto, Ontario, M5S 3G4, Canada}
\email{hklo@comm.utoronto.ca}

\date{\today}
\begin{abstract}
The noisy-storage model allows the implementation of secure two-party protocols under the sole assumption
that no large-scale reliable quantum storage is available to the cheating party. 
No quantum storage is thereby required for the honest parties.
Examples of such protocols
include bit commitment, oblivious transfer and secure identification. 
Here, we provide a guideline for
the practical implementation of such protocols. In particular, we 
analyze security in a practical setting where the honest parties themselves are unable to perform perfect
operations and need to deal with practical problems such as errors during transmission and detector inefficiencies.
We provide explicit security parameters for two different experimental setups using weak coherent, and parametric
down conversion sources. In addition, we analyze a modification of the protocols based on decoy states.
\end{abstract}
\maketitle

\tableofcontents
\newpage
\section{Introduction}

\begin{figure}[h]
\begin{center}
\includegraphics{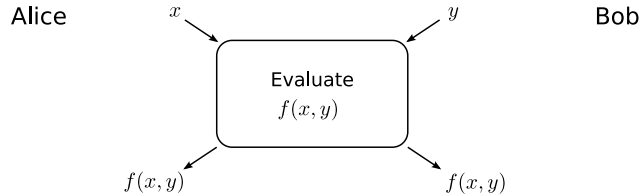}
\caption{Alice holds an input $x$ (e.g.\ her password), and Bob holds
  an input $y$ (e.g.\ the password an honest Alice should possess), and
  they want to obtain the value of some function $f(x,y)$ (e.g.\ the
  equality function).} \label{fig:SFE}
\end{center}
\end{figure}

Quantum cryptography allows us to solve cryptographic tasks without resorting to unproven computational assumptions. 
One example is quantum key distribution (QKD) which is well-studied within quantum information \cite{qkd1,qkd2}.
In QKD, the sender (Alice) and the receiver (Bob) trust each other, but want to shield their 
communication from the prying eyes of an eavesdropper. In many other cryptographic problems, however,
Alice and Bob themselves do \emph{not} trust each other, but nevertheless want to 
cooperate to solve a certain task. An important example of such a task is secure identification. Here, Alice wants to identify herself to Bob
(possibly an ATM machine) without revealing her password.
More generally, Alice and Bob wish to perform \emph{secure function
  evaluation} as depicted in Figure~\ref{fig:SFE}.

In this scenario, security means that the legitimate users should not
learn anything beyond this specification. That is, Alice should not
learn anything about $y$ and Bob should not learn anything about $x$,
other than what they may be able to infer from the value of $f(x,y)$.
Classically, it is possible to solve this task if one is willing to
make computational assumptions, such as that factoring of large
integers is difficult. Sadly, these assumptions remain unproven.
Unfortunately, even quantum mechanics does not allow us to implement
such interesting cryptographic primitives without further
assumptions~\cite{lo:insecurity,mayers:trouble,lo&chau:bitcom2,lo&chau:bitcom,mayers:bitcom}.

\subsection{The noisy-storage model}

\begin{figure}
\begin{center}
\includegraphics{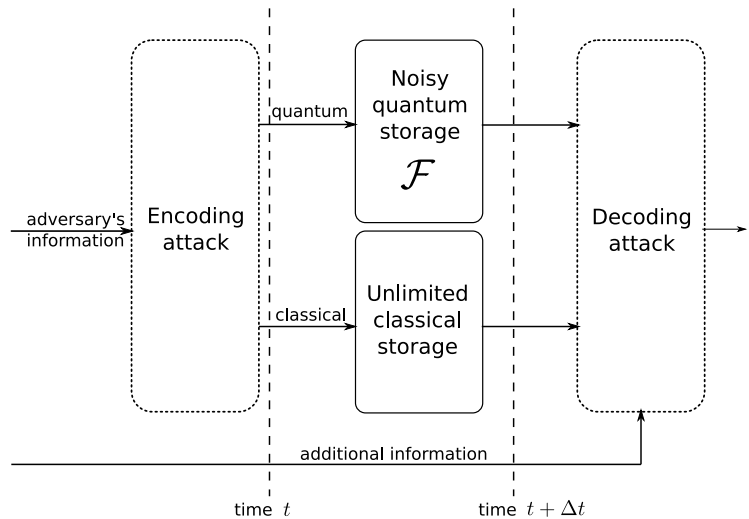}
\caption{During waiting times $\Delta t$, the adversary must use his
  noisy-quantum storage described by the CPTP map $\cF$. Before using
  his quantum storage, he performs any (error-free) ``encoding
  attack'' of his choosing, which consists of a measurement or an
  encoding into an error-correcting code. After time $\Delta t$, he
  receives some additional information that he can use for
  decoding.} \label{fig:noisystorage}
\end{center}
\end{figure}

The noisy-storage model (NSM) allows us to obtain secure two-party protocols under the \emph{physical} assumption that any cheating
party does not posses a large reliable quantum storage. First introduced in~\cite{prl:noisy,steph:diss}, the NSM
has recently~\cite{noisy:new} 
been shown to encompass both the case where the adversary has a bounded amount of noise-free 
storage~\cite{serge:bounded,serge:new} (also known as 
the bounded-storage model), as well as the case where the adversary 
has access to a potentially large amount of noisy storage.
This last assumption is well justified given the state of present day technology, and the fact that merely transferring
the state of a photonic qubit onto a different carrier (such as an atomic ensemble) is typically already noisy, even if the
resulting quantum memory is perfect. In the
protocols considered, the honest parties themselves do not require any quantum storage
at all.
We briefly review the NSM here for completeness. 
Without loss of generality, noisy quantum storage is described by a family of completely positive 
trace-preserving maps $\{\cF_t:\cB(\cH_{in})\rightarrow\cB(\cH_{out})\}_{t>0}$, where $t$ is the time that the adversary 
uses his storage device. 
An input state $\rho$ on~$\cH_{in}$ stored at time~$t_0=0$ decoheres over time, resulting in a state $\cF_{t}(\rho)$ of the 
memory at time~$t$. 
We make the minimal assumption that the noise is Markovian, meaning that the adversary does not gain any advantage by delaying 
the readout whenever he wants to
retrieve encoded information: waiting longer only degrades the information further.
The only assumption underlying the noisy-storage model consists 
in demanding that the adversary can only keep quantum information in this noisy storage device. In particular, he is otherwise 
completely unrestricted -- for example, he can perform arbitrary (instantaneous) quantum computations 
using information from the storage device and additional ancillas. 
In particular, he is able to perform perfect, noise-free, quantum computation and communication.
However, after his computation he needs to discard all quantum information except 
what is contained in the storage device, where he may prepare an arbitrary encoded state on~$\cH_{in}$.
This scenario is illustrated in Figure~\ref{fig:noisystorage}.

How can we obtain security from such a physical assumption? We consider protocols which force the adversary to store quantum 
information for extended periods to gain information: This is achieved by using certain time delays~$\Delta t$ at specific points 
in the protocol (e.g., before starting a round of communication). This 
forces the adversary to use his device for a time at least~$\Delta t$ if he wants to preserve quantum information.
Due to the Markovian assumption, it suffices to analyze security 
for the channel $\cF=\cF_{\Delta t}$.  
Hence the security model can be summarized as follows:
\begin{itemize}
\item
The adversary has unlimited classical storage, and (quantum) computational resources. He is able to perform any operations
noise-free and has access to a noise-free quantum channel.
\item
Whenever the protocol requires the adversary to wait for a time $\Delta t$, he has to measure/discard all his quantum information except what he can encode (arbitrarily) into $\cH_{in}$. This information then undergoes noise described by~$\cF$.
\end{itemize}

We stress that in contrast to the adversary's potential resources allowed in this model, the technological demands on honest 
parties are minimal: in our protocol, honest parties merely need to prepare and measure BB84-encoded qubits~\footnote{That is, qubits encoded in one
of two conjugate bases, such as the computational and Hadamard basis.} and do not require 
any quantum storage. 

\subsection{Challenges in a practical implementation}
In this work we focus on how to put the protocols
of~\cite{noisy:new} into practice. 
Unfortunately, the theoretical analysis of~\cite{noisy:new} assumes perfect single-photon sources that
are not available yet \cite{sp1,sp2}. Here, we remove this assumption leading to a slightly modified
protocol that can be implemented immediately using today's technology.
At first glance, it may appear
that the security analysis for a practical implementation differs
little from the problems encountered in practical realizations of
QKD. After all, the quantum communication part of the protocols
in~\cite{noisy:new} consists of Alice sending BB84 states to Bob. Yet,
since now the legitimate users do not trust each other, the analysis
differs from QKD in several fundamental aspects. Intuitively, these
differences arise because Alice and Bob do not cooperate to check on
an outside eavesdropper. Quite on the contrary, Alice can never rely
on anything that Bob says. A second important aspect that
differentiates the setting in~\cite{noisy:new} from QKD lies in the
task the cryptographic protocols aim to solve.  For instance, secure
identification is particularly interesting at extremely short
distances, for which Alice would ideally use a small, low power,
portable device. Bob, on the other hand, may use more bulky
detectors. At such short distances, we could furthermore use visible
light for which much better detectors exist than those typically used
in QKD at telecom wavelengths.
It is an interesting experimental challenge to come up with suitable
devices. Small handheld setups have been proposed to perform QKD at short distance~\cite{rarity:small},
which we can also hope to use here. The QKD devices of~\cite{rarity:small} have been devised to distribute
non-reusable authentication keys which could also be employed for identification. 
At such short distance, this could also be achieved by for example loading keys onto a USB stick at a trusted loading station at
a bank for instance.  We emphasize that our work is in spirit very different in that we allow
authentication keys to be reused over and over again, just as traditional passwords~\cite{DFSS07}.

We first analyze a generic experimental setup in
Section~\ref{sec:abstractSource}.  More specifically, we present a
source-independent characterization of such a setup and discuss all
parameters that are necessary to evaluate security in the NSM.
Especially important is that in any real-world setting even the honest
parties do not have access to perfect quantum operations, and the
channel connecting Alice and Bob is usually noisy. The challenge we
face is to enable the honest parties to execute the protocol
successfully in the presence of errors, while ensuring that the
protocol remains secure against any cheating party. We shall always
assume a worst-case scenario where a cheating party is able to perform
perfect quantum operations and does not experience channel noise, its
only restriction is its noisy quantum storage.

The primary source of errors at short distances lies in the low
detector efficiencies of present day single-photon detectors. For telecom
wavelengths these detector efficiencies $\eta_D$ lie at roughly $10\%$, where
at visible wavelengths one can use detectors of about $70\%$
efficiency. Hence, a considerable part of the transmissions will be
lost. In Section~\ref{sec:wseErrors}, we augment the protocol for weak
string erasure
presented in~\cite{noisy:new} to deal with such erasure errors. 
This protocol is the main ingredient to realize the primitive of oblivious transfer, which can be used to solve the 
problem of computing a function $f(x,y)$. 
The second source of errors lies in bit errors which result from noise on the channel itself or imperfections in Alice
and Bob's measurement apparatus. At short distances, such errors will typically be quite small. 
In Section~\ref{sec:otFromWSEE}, we show how to augment the protocol for oblivious transfer to deal with bit errors.
It should be noted that we treat these errors in the classical 
communication part of the protocols, independently of erasure errors,
and similar techniques may be used in other schemes based on weak string erasure in the future.

To obtain security, we have to make a reasonable estimation of the
errors that the honest parties expect to occur.  We state the
necessary parameters in Section~\ref{sec:abstractSource} and provide
concrete estimates for two experimental setups in
Section~\ref{sec:concreteSource}. In particular, we present explicit
security parameters for a source of weak coherent pulses, and a
parametric down conversion (PDC) source.  Throughout, we assume that
the reader is familiar with commonly used entropic quantities also
relevant for QKD, and quantum information. An introduction to all
concepts relevant for security in the NSM is given
in~\cite{noisy:new}. 

\section{General setup}\label{sec:abstractSource}\label{sec:parameters}

\begin{figure}
\begin{center}
\includegraphics{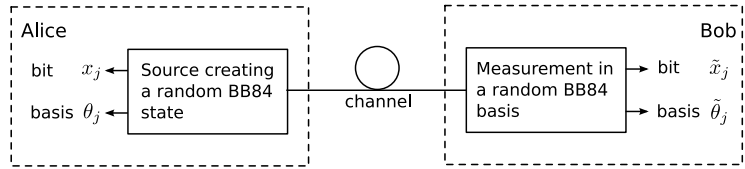}
\caption{A general setup for weak string erasure.}
\label{fig:general}
\end{center}
\end{figure}

Before turning to the actual protocols, we need to investigate the parameters involved in an
experimental setup. 
The quantum communication part of all the protocols in the NSM is a simple 
scheme for weak string erasure
which we will describe in detail in the next section. In each round of this protocol, Alice 
chooses one of the four possible BB84 states~\cite{bb84} at random and sends this state to Bob.
Bob now measures randomly the state received either in the computational or in the Hadamard basis. 
Such a setup is characterized by a source held by Alice, and a measurement apparatus held by Bob 
as depicted in Figure~\ref{fig:general}. 
The source can as well include a measurement device, depending 
on the actual state preparation process (e.g. when a PDC source acts as a triggered 
single-photon source). 
If Alice is honest, we can trust the source entirely, 
which means that in principle we have full knowledge of its parameters. 
Note, however, that in any practical setting the parameters of the source will undergo small fluctuations. 
For clarity of exposition, we do not take these fluctuations into account explicitly, but assume
that all the parameters below are worst-case estimates of what we can reasonably expect from our source.

\subsection{Source parameters}

Unfortunately, we do not have access to a perfect single-photon source in a practical setting \cite{sp1,sp2}, but can only arrange
the source to emit a certain number of photons with a certain probability. To approximate a single-photon source, 
we will later let Alice perform some measurements herself to exclude multi-photon events in the case of a PDC source.
The following table
summarizes the two relevant probabilities we need to know in any implementation.
When using decoy states, we will frequently add an index $s$ to all parameters to specify a particular source $s$
that is used.
\bigskip
\begin{center}
\begin{tabular}{|l|l|}
\hline
probability & description\\
\hline
$p^n_\src$ & the source emits $n$ photons.\\[2pt]
\hline
$p^{n}_\sent$ & the source emits $n$ photons conditioned on the
event\\
& that Alice concludes that \emph{one} photon has been 
emitted.\\
\hline
\end{tabular}
\end{center}
\smallskip
In our analysis, we will be interested in bounding the number of
single-photon emissions in $M$ rounds of the protocol, which can be
achieved using the well-known Chernoff's inequality (see
e.g.~\cite{AS00}): Suppose we have a source that emits a single photon
with probability $p^1_\src$ and a different number of photons
otherwise.  How many single-photon emissions do we expect?
Intuitively, it is clear that in $M$ rounds we have roughly $p^1_\src
M$ many. Yet, in the following we need to consider a small interval
around $p^1_\src M$, such that the probability that we do not fall
into this interval is extremely small.  More precisely, we want that
\begin{align} \label{eq:chernoff}
\Pr[|S - p^1_\src M| \geq \zeta_\src^1 M] \leq \eps,
\end{align}
where $S$ is the number of single-photon emissions.
To apply Chernoff's inequality, let $X_j=1$ denote the event where a single-photon emission occurred, and 
let $X_j = 0$ otherwise, giving us $S = \sum_j X_j$. We then demand that
\begin{align}
2 e^{- 2 (\zeta_\src^1)^2 M} \leq \eps,
\end{align}
which can be achieved by choosing $\zeta_\src^1 = \sqrt{\ln(2/\eps)/(2M)}$. Operationally this means that the number
of single-photon emissions lies in the interval $[(p^1_\src - \zeta^1_\src)M,(p^1_\src + \zeta_\src^1)M]$, except
with probability $\eps$. Note that for $M$ being very large we indeed
have $\zeta^1_\src \approx 0$, leaving us with approximately $p^1_\src M$ many single-photon emissions. 
By exactly the same argument, if now $M$ refers to the number of rounds in the protocol where Alice concluded
the source emitted single-photons, the actual number of single-photon rounds within these post-selected events
lies in the interval $[(p^1_\sent - \zeta^1_\sent)M, (p^1_\sent + \zeta^1_\sent)M]$ for 
$\zeta^1_\sent = \sqrt{\ln(2/\eps)/(2M)}$, except with probability $\eps$. 
We will make use of this argument repeatedly and use $\zeta^x_y$ to denote the interval when considering an event
that occurs with probabiity $p^x_y$.

We would like to emphasize that for our security proof to work, we
only need a conservative \emph{lower bound} on the number of
single-photon emissions. Should there be some intensity
fluctuations in Alice's laser provided that we know the \emph{worst} case
(i.e., a conservative lower bound $p^1_{\src}$) in the asymptotic case
of large $M$, then the discussion for the finite-size case will go
through if we consider a \emph{one-sided} bound in
Equation~\eqref{eq:chernoff}. i.e., 
$\Pr[S \leq ( p^1_{\src} - \zeta^1_{\src} ) M ] \leq \eps$.

\subsection{Error parameters}

For any setup, we need to determine the following error parameters. 
These parameters should be a reasonable
estimate that is made once for a particular experimental implementation and fixed during subsequent executions of the protocol.
For instance, for a given device meant to be used for identification, these estimates would be fixed during construction.

\subsubsection{Losses}
As mentioned above, the primary restriction in a practical setting arises from
the loss of signals. These losses can occur on the channel, or be
caused by detector inefficiencies. The following table summarizes all the 
probabilities we need.  Throughout, we use the superscripts $h$ and
$d$ to indicate that these parameters apply to an honest or dishonest
party respectively. 
\bigskip
\begin{center}
\begin{tabular}{|l|l|}
\hline
probability & description\\
\hline
$p^n_\lossC$ & $n$ photons are erased on the channel\\[2pt]
\hline
$p^h_\clickB$ & honest Bob observes a click in his detection apparatus\\[2pt]
\hline
$p^h_\lostB$ & honest Bob observes no click in his detection apparatus\\[2pt]
\hline
$p^{h|n}_\clickB$ & honest Bob observes a click in his detection apparatus,\\
& conditioned on the event that Alice 
sent $n$ photons.\\[2pt]
\hline
$p^h_\slostB$ & honest Bob observes no click from the signal alone\\[2pt]
\hline
$\pdark$ & an honest player obtains a click when the signal was a vacuum state (dark count)\\[2pt]
\hline
\end{tabular}
\end{center}
\smallskip
\noindent
Note that we have
\begin{align}
p^h_\lostB = \sum_{n=0}^{\infty} p^n_\src \, p^{h|n}_\lostB\ ,
\end{align}
and again the number of rounds we expect to be lost can be bounded to lie in the interval 
$[(p^h_\lostB - \zeta^h_\lostB)M,(p^h_\lostB + \zeta^h_\lostB)M]$ with $\zeta^h_\lostB = \sqrt{\ln(2/\eps)/(2M)}$, except
with probability $\eps$.

\subsubsection{Bit errors}\label{sec:bitErrors}

The second source of errors are bit-flip errors that can occur due to imperfections in Alice's or Bob's measurement apparatus
or due to noise on the channel. We use the following notation for the probability of such an event in the case that Bob is
honest. This probability depends on the detection error $e_{\rm det}$
in our experimental setup, i.e., on the probability that a 
signal sent by Alice produces a click in 
the erroneous detector on Bob's side, and on $\pdark$. The quantity $e_{\rm det}$
characterizes the alignment and stability of the optical system. 
\bigskip
\begin{center}
\begin{tabular}{|l|l|}
\hline
parameter & description\\
\hline
$e_{\rm det}$ & detection error\\[2pt]
\hline
$p^h_\errB$ & honest Bob outputs the wrong bit\\[2pt]
\hline
\end{tabular}
\end{center}
\smallskip
\noindent
For a single bit $b \in \01$, a bit-flip error is described by the classical binary symmetric channel with error parameter $p_{\rm err}$
\begin{align}
\cS_{p_{\rm err}}(b) = \left\{  \begin{array}{c@{\quad}l} 
b & \mbox{ with probability $1-p_{\rm err}$\, , }\\
(1-b) & \mbox{ with probability $p_{\rm err}$ .} \end{array} \right.
\end{align}
When each bit of a $k$-bit string is independently affected by
bit-flip errors, the noise can be described by the channel
\begin{align}\label{eq:bitErrors}
\cE_{p_{\rm err}} = \cS_{p_{\rm err}}^{\otimes k}\ ,
\end{align}
where we omit the explicit reference to $k$ on the l.h.s. when it is clear from the context.

\subsection{Parameters for dishonest Bob}

Recall our conservative assumption that a dishonest party is only
restricted by its noisy quantum storage, but can otherwise perform
perfect quantum operations and has access to a perfect channel.  Yet,
even for a dishonest Bob there are some errors he cannot avoid, caused
by the imperfections in Alice's apparatus.  If Alice's source simply
outputs no photon for example, then even a dishonest Bob cannot detect
the transmission which is captured by the following parameter.
\bigskip
\begin{center}
\begin{tabular}{|l|l|}
\hline
probability & description\\
\hline
$p^d_\lostB$ & dishonest Bob observes no click in his detection apparatus\\[2pt]
\hline
\end{tabular}
\end{center}
\smallskip
\noindent
Generally, we have $p^d_\lostB = p^0_\sent$. 
In the protocols that follow, we will ask an honest Bob to report any round as missing that has not resulted in a click. 
Without loss of generality, we can assume that even a dishonest Bob
will report a particular round as lost when he does not observe a
click. Of course, if Bob is dishonest he potentially chooses to report additional rounds as missing. 

In our analysis, we also have to evaluate the following probability which depends on the experimental setup, as well
as on our choice of protocol parameters. 
\bigskip
\begin{center}
\begin{tabular}{|l|l|}
\hline
probability & description\\
\hline
$p^{d,n}_\errB$ & dishonest Bob outputs the wrong bit if Alice sent $n$ photons,\\
& and he gets the basis information for free\\[2pt]
\hline
\end{tabular}
\end{center}
\smallskip
\noindent
 
\section{Weak string erasure with errors}\label{sec:wseErrors}\label{sec:WSEEnoDecoy}

The basic quantum primitive upon which all other protocols
in~\cite{noisy:new} are based is called weak string erasure.
Intuitively, weak string erasure provides Alice with a random $m$-bit
string $X^m$ and Bob with a random set of indices $\cI \in 2^{[m]}$
and the substring $X_{\cI}$ of $X^m$ restricted to the elements in
$\cI$~\footnote{We use $2^{[m]}$ to denote all subsets of the set $[m] = \{1,\ldots,m\}$}. 
If Bob is honest, then we demand that whatever attack
dishonest Alice mounts, she cannot gain any
information about which bits Bob has learned. That is, she cannot gain
any information about $\cI$.  If Alice herself is honest, we demand
that the amount of information that Bob can gain about the string
$X^m$ is limited.

\begin{figure}
\begin{center}
\includegraphics{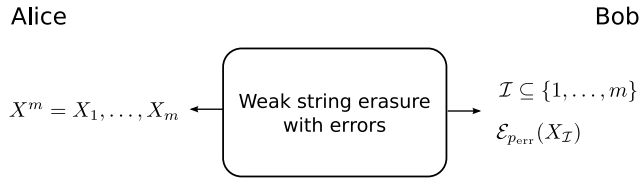}
\end{center}
\caption{Weak string erasure with errors when both parties are honest. $\cE_{p_{\rm err}}$ denotes the bit-error channel
defined in~\eqref{eq:bitErrors}.}
\label{fig:WSEE}
\end{figure}

We now present an augmented version of the weak string erasure protocol proposed in~\cite{noisy:new} 
that allows us to deal with the inevitable errors 
encountered during a practical implementation. We thereby address the two possible errors separately: losses
are dealt with directly in weak string erasure. Bit-flip errors, however, are not corrected in weak string erasure itself,
but in subsequent protocols \footnote{Subsequent protocols will use only part of the string $X^m$, and hence
allow us to decrease the amount of error-correcting information needed.}.
We will thus implement \emph{weak string erasure with errors} where the substring $X_{\cI}$ is allowed to be affected by bit-flip errors. That is, honest Bob actually receives $\cE_{p_{\rm err}}(X_{\cI})$ where $\cE_{p_{\rm err}}$ is the classical channel corresponding to the bit errors as given in~\eqref{eq:bitErrors}, with $k = |\cI|$ being the length 
of the string $X_{\cI}$. Figure~\ref{fig:WSEE} provides an intuitive description
of this task.

We now provide an informal definition of weak string erasure with errors. A formal definition can be found in Appendix~\ref{app:wsee}.
Even in this informal definition we need to quantify the knowledge that a cheating Bob has about the string $X^m$
given access to his entire system $B'$~\footnote{We use $B'$ to differentiate it from the system $B$ an honest Bob holds.}.
This quantity has a simple interpretation in terms of the min-entropy as $\hmin(X^m|B') = - \log P_{\rm guess}(X^m|B')$, 
where $P_{\rm guess}(X^m|B')$ represents the probability that Bob guesses $X^m$, maximized over all
measurements of the quantum part $B'$. The quantity $\hmin^\varepsilon(X^m|B')$ thereby behaves like $\hmin(X^m|B')$, except with probability
$\varepsilon$. We refer to~\cite{noisy:new} for an 
introduction to these quantities and their use in the NSM.

\begin{definition}[Informal]
An {\em $(m,\lambda,\varepsilon,p_{\rm err})$-weak string erasure protocol with errors} (WSEE) is a protocol 
between Alice and Bob satisfying the following properties, where $\cE_{p_{\rm err}}$ is defined as in~\eqref{eq:bitErrors}:
\begin{description}
\item[Correctness:]
If both parties are honest, then Alice obtains a randomly chosen $m$-bit string $X^m \in \{0,1\}^m$, 
and Bob obtains a randomly chosen subset $\cI \subseteq [m]$, as well as the string $\cE_{p_{\rm err}}(X_{\cI})$.
\item[Security for Alice:] If Alice is honest, then the amount of information Bob has about $X^m$ is limited
to
\begin{align}
\frac{1}{m}\hmin^\varepsilon(X^m|B')\geq \lambda\ ,
\end{align}
where $B'$ denotes the total state of Bob's system.
\item[Security for Bob:] If Bob is honest, then Alice learns nothing about $\cI$.
\end{description}
\end{definition}

We are now ready to state a simple protocol for WSEE. 
We thereby introduce
explicit time slots into the protocol. If Alice herself concludes that no photon or a multi-photon
has been emitted in a particular time slot, she simply discards this round and tells
Bob to discard this round as well. Since this action represents no security problem for us, we will
for simplicity omit these rounds all-together when stating the protocol below.
This means that the number of rounds $M$ in the protocol below, actually refers to the set of
post-selected pulses that Alice did count as a valid round.

In addition, introducing time slots enables Bob to report a
particular bit as missing, if he has obtained no click in a particular
time slot. Alice and Bob will subsequently discard all missing rounds.
This does pose a potential security risk, which we need to analyze and hence we explicitly include this
step in the protocol below.

\begin{protocol}{1}{Weak String Erasure with Errors (WSEE)}
{Outputs: $x^m \in \sbin^m$ to Alice, 
$(\cI,z^{|\cI|}) \in 2^{[m]} \times \sbin^{|\cI|}$ 
to Bob.}\label{proto:wse}
\item[1.] {\bf Alice:} Chooses a string $x^M \in_R \01^M$ and basis-specifying 
string $\theta^M \in_R \01^M$ uniformly at random. 
\item[2.] {\bf Bob: } Chooses a basis string $\tilde{\theta}^M \in_R \01^M$ uniformly at random. 
\item[3.] In time slot $i=1,\ldots,M$ (considered a valid round by Alice): 
\begin{enumerate}
\item {\bf Alice:}
Encodes bit $x_i$ in the basis given by $\theta_i$ (i.e., as $H^{\theta_i}\ket{x_i}$), and sends
the resulting state to Bob.
\item {\bf Bob:} Measures in the basis
given by $\tilde{\theta}_i$ to obtain outcome $\tilde{x}_i$.
If Bob obtains no click in this time slot, he records round $i$ as missing.
\end{enumerate}

\item[4.] {\bf Bob: } Reports to Alice which rounds were missing.

\item[5.] {\bf Alice: } If the number of rounds that Bob reported missing does not lie in the 
interval $[(p^h_\lostB - \zeta^h_\lostB)M,(p^h_\lostB + \zeta^h_\lostB)M]$, then Alice aborts 
the protocol. Otherwise, she deletes all bits from $x^M$ that Bob reported missing. Let $x^m \in \01^m$
denote the remaining bit string, and let $\theta^m$ be the basis-specifying string for the remaining rounds.
Let $\tilde{\theta}^m$, and $\tilde{x}^m$ be the corresponding strings for Bob.

\item[Both parties wait time $\Delta t$.]

\item[6.] {\bf Alice: } Sends the basis information $\theta^m$ to Bob, and outputs $x^m$.
\item[7.] {\bf Bob: } Computes $\cI \assign \{i \in [m] \mid \theta_i = \tilde{\theta}_i\}$, and outputs $(\cI,z^{|\cI|}):=(\cI,\tilde{x}
_{\cI})$.
\end{protocol}

\subsection{Security analysis}

\subsubsection{Parameters}
We prove the security of Protocol 1 in Appendix~\ref{app:wsee}, where
our analysis forms an extension of the proof presented
in~\cite{noisy:new}.  The security proof for dishonest Alice is analogous to~\cite{noisy:new}.
The only novelty is to ensure that allowing Bob
to report rounds as missing does not compromise the security.  Here,
we focus on weak string erasure with errors, when the adversary's
storage is of the form $\cF = \cN^{\otimes \nu M_{\rm store}}$, and
$\cN$ obeys the strong converse property \cite{rs:converse}. An important example is 
the $d$-dimensional depolarizing channel. For this case, we can give
explicit security parameters in terms of the amount of noise generated
by $\cN$.  The quantity $\nu$ denotes the storage rate, and $M_{\rm
  store}$ is the number of single-photon emissions that we expect an
honest Bob to receive for large $M$.  That is 
\begin{align}\label{mstore}
M_{\rm store} := p^{1}_{\rm sent} \cdot p^{h|1}_{\clickB} \cdot M\ .
\end{align}
We hence allow Bob's storage size to be determined as in the idealized setting
of~\cite{noisy:new}, where we have only single-photon emissions.
Throughout, we let $M^{(n)}$ denote the number of $n$ photon emissions
in $M$ valid rounds, and use $r^{(n)}$ to denote the fraction of these $n$-photon
pulses that Bob decides to report as missing. Clearly,
$r^{(n)}$ is not a parameter we can evaluate, but depends on the
strategy of dishonest Bob.  Finally, we use $M^{(n)}_{\rm left} =
(1-r^{(n)}) M^{(n)}$ to denote the number of $n$-photon pulses that
are left. Note that $M^{(n)}_{\rm left}$ is a function of $r^{(n)}$
chosen by Bob according to certain constraints which we investigate
later.
A proof of Theorem~\ref{thm:WSEEmaintext}, as well as a
generalization to other channels $\cF$ not necessarily of the form
$\cF = \cN^{\otimes \nu M_{\rm store}}$, can be found in
Appendix~\ref{app:wsee}. Here, we state the theorem for a worst-case
setting which can be obtained using~\eqref{eq:boundRate}. This result
is independent of the actual choice of signals that Bob chooses to
report as missing.  For simplicity, we present the theorem omitting
terms that vanish for large $M$.  These terms are, however, considered
in Appendix~\ref{app:wsee}.

\begin{theorem}[WSEE]\label{thm:WSEEmaintext}
  Let Bob's storage be given by $\cF=\cN^{\otimes \nu M_{\rm store}}$
  for a storage rate~$\nu>0$, $\cN$ satisfying the strong converse
  property~\cite{rs:converse} and having capacity~$C_\cN$ bounded by
\begin{align}
  C_\cN\cdot\nu< \left(\frac{1}{2} - \delta\right) \frac{p^1_{\rm
      sent} - p^h_\lostB + p^d_\lostB}{p^1_{\rm sent} \cdot
    p^{h|1}_\clickB} \, .
\end{align} 
Then Protocol~1 is an $(m,\lambda(\delta),\eps(\delta),p^h_{\errB})$-weak string erasure protocol with errors 
with the following parameters:
Let $\delta\in ]0,\frac{1}{2}-C_\cN\cdot \nu[$.
Then the min-entropy rate $\lambda(\delta)$ is given by 
\begin{align}
\lambda(\delta) &=
\min_{\{r^{(n)}\}_n}
\frac{1}{m} 
\left[
\nu \cdot \gamma^\cN\left(\frac{R}{\nu}\right) \cdot M_{\rm store} - 
\sum_{n=2}^{\infty} M^{(n)}_{\rm left} \log\left(1 - p^{d,n}_{\errB}\right)\right]\ ,
\end{align}
where $\gamma^\cN$ is the strong converse
parameter of $\cN$ (see~(\ref{eq:converseParameter})) and 
the minimization is taken over all $\{r^{(n)}\}_n$ such that 
$\sum_{n=1}^{\infty} r^{(n)} M^{(n)} \leq M^d_{\rm report}$, $M_{\rm
  store}$ is given by (\ref{mstore}), and

\bigskip
\begin{tabular}{ll}
$m = \sum_{n=1}^{\infty} M^{(n)}_{\rm left}$ &
(the number of remaining rounds) 
\ ,\\[2mm]
$M^d_{\rm report} = (p^h_\lostB - p^d_\lostB)M$&
(the number of rounds dishonest Bob can report missing)   
\ ,\\[2mm]
$R  = \left(\frac{1}{2} - \delta\right) \frac{1-r^{(1)}}{p^{h|1}_{\clickB}}$&
(the rate at which dishonest Bob has to send information through storage)
\ ,
\end{tabular}
\bigskip

\noindent
for sufficiently large $M$.
The error has the form
\begin{align}
\eps(\delta) \leq 4 \exp\left( 
- \frac{\delta^2}{512(4 + \log\frac{1}{\delta})^2} \cdot 
(p^1_\sent - p^h_\lostB + p^d_\lostB ) M  
\right)\ .
\end{align}
\end{theorem}

What kind of channels $\cN: \bop(\hin) \rightarrow \bop(\hout)$
satisfy the strong converse property? It was
recently shown in~\cite{rs:converse} that all channels for which the maximum $\alpha$-norm
is multiplicative, and which are group covariant, that is $\cN(g \rho
g^\dagger) = g \cN(\rho) g^\dagger$ for all $g \in G$ where $g$ acts
irreducibly on the output space $\hout$, satisfy this property. An
important example of such a channel is the $d$-dimensional
depolarizing channel given as
\begin{align}\label{eq:depol}
\cN_r(\rho) := r \rho + (1-r)\frac{\id}{d}\ ,
\end{align}
which replaces the input state $\rho$ with the completely mixed state $\id/d$ with probability $1-r$.
Security parameters for this channel 
can be found in~\cite{noisy:new}
for the case of a 
perfect setup with a single-photon source, assuming no errors nor detection inefficiencies.

\subsubsection{Limits to security}
Before analyzing in detail concrete practical implementations based on a weak coherent source, and a PDC source, 
we investigate when security can be obtained at all 
for the $d$-dimensional
depolarizing channel
as a function of $p^1_{\rm sent}$, $p^d_\lostB$, $p^h_\lostB$, and
$p^{h|1}_\lostB$ in comparison to the storage parameters $r$ and $\nu$.
Note that for the security parameter $\eps(\delta)$ to vanish we need
\begin{align}\label{eq:cond1}
p^1_{\rm sent} - p^h_\lostB + p^d_\lostB > 0\ .
\end{align}
Second, we require (in the limit of large $M$ where we may choose $\delta \rightarrow 0$) that
\begin{align}\label{eq:cond2}
C_{\cN_r} \cdot\nu< \frac{1}{2} 
\frac{p^1_{\rm sent} - p^h_\lostB + p^d_\lostB}{p^1_{\rm sent} \cdot p^{h|1}_\clickB}\ ,
\end{align} 
where $C_{\cN_r}$ is given by~\cite{king:depol}
\begin{align}
C_{\cN_r} = \log d + \left(r + \frac{1-r}{d}\right) \log \left(r + \frac{1-r}{d}\right)
+ (d-1) \frac{1-r}{d} \log \frac{1-r}{d}\ .
\end{align}
In Sections~\ref{sec:wcp} and~\ref{sec:pdc} we provide sample trade-offs between $r$ and $\nu$ for some typical values
of the source parameters, and the losses.

To determine the magnitude of the actual security parameters, we need
to evaluate the strong converse parameter $\gamma^{\cN}$~\cite{rs:converse}. In the case of the $d$-dimensional depolarizing channel it can be expressed as~\cite{noisy:new}
\begin{align}\label{eq:converseParameter}
\gamma^{\cN}(\hat{R}) := \max_{\alpha \geq 1} \frac{\alpha-1}{\alpha}
\left\{\hat{R} - \log d + \frac{1}{1-\alpha} \log \left[\left(r + \frac{1-r}{d}\right)^\alpha 
+ (d-1)\left(\frac{1-r}{d}\right)^\alpha \right] \right\}\ .
\end{align}
For a general definition and discussion on how to evaluate this parameter for other channels
see \cite{rs:converse,noisy:new}.
For simplicity, we consider here
a setup where Bob always gains full information from a multi-photon emission, that is
$p^{d,n}_{\errB} = 0$ for $n>1$. This means that he will never report any such rounds as missing, 
that is, $r^{(n)} = 0$ for $n > 1$. From~\eqref{eq:boundRate} it follows that
\begin{align}\label{eq:lambdaOptimize}
\lambda(\delta) \geq \frac{1}{m} \left\{ \nu \cdot \gamma^{\cN}\left[\frac{1}{\nu}\left(\frac{1}{2}-\delta\right) 
\frac{p^1_{\rm sent} - p^h_\lostB + p^d_\lostB}{p^1_{\rm sent} \cdot p^{h|1}_\clickB}\right]\right\} M_{\rm store}\ ,
\end{align}
providing the security conditions~\eqref{eq:cond1} and~\eqref{eq:cond2} are satisfied.
In Sections~\ref{sec:wcp} and~\ref{sec:pdc} we plot $\lambda(\delta)$ for a variety of parameter choices
for a weak coherent and a PDC source respectively.

\subsection{Using decoy states}\label{sec:decoy}\label{sec:proofDecoy}

We now consider a slight modification of the protocol above, where we make use of so-called decoy states as they are also
used in QKD~\cite{decoy,decoy1,decoy2}.
The main idea consists of Alice randomly 
choosing 
a particular setting of her photon source according to a distribution $P_S$ over
some set of settings $\cS$
for each state she sends to Bob. One of these settings (signal setting)
corresponds to the configuration of the source she would normally use to
execute the weak string erasure protocol above, all others (decoy settings) are used to test the behavior of \emph{dishonest} Bob.
In our setting, the effect of using decoy states is that dishonest Bob needs to behave roughly the same as honest Bob
when it comes to choosing which rounds to report as missing.
This enables us to place a better bound on the parameter $r^{(1)}$, which can lead to a significant increase in the
set of detection efficiencies for which we can hope to show security (e.g., for a weak coherent source see Section~\ref{sec:wcpDecoy}),
and translates into an enhancement of the rate $R$ given by~\eqref{eq:defR} and~\eqref{eq:boundRate} at which
the adversary needs to transmit information through his storage, if he wants to break the security of the protocol.

We briefly describe how we make use of decoy states, before turning to the actual protocol.
For each source setting, Alice can compute the \emph{gain}, that is, the probability that Bob observes a click. 
Here we consider only the number of rounds $M$ which Alice determines to be valid, and all probabilities are as explained
in Section~\ref{sec:parameters} conditioned on the event that Alice declared the round to be valid.
We can then write the gain of honest Bob when Alice uses setting $s$, averaged over all possible numbers of photons, as
\begin{align}
Q_s^h = p^h_\clickBs = \sum_{n=0}^{\infty} p^n_\sents p^{h|n}_\clickB\ .
\end{align}
Note that $p^{h|n}_\clickB$ thereby does \emph{not} depend on the source setting $s$, 
even though Bob \emph{can} gain information about the setting $s$ by making a photon number measurement, since not 
all photon numbers are equally likely to occur for the different settings. Yet, since the photon number is the only information
that Bob obtains, we can without loss of generality assume that his strategy is deterministic
and depends only on the observed photon number. 
By counting the number of rounds that Bob reports missing, Alice obtains an estimate of this gain as
\begin{align}
Q_s^{\rm meas} = \frac{M_{{\rm left},s}}{M_s}\ .
\end{align}
The parameter $M_s$ denotes the number of valid rounds in which Alice uses setting $s$, and $M_{{\rm left},s}$ represents the number of such rounds
that Bob did not report as missing. For an honest Bob, we have $Q_s^{\rm meas} \approx Q_s^h$ in the limit of large $M_s$.
For finite $M_s$, we conclude that $M_{{\rm left},s}$ lies in the interval 
$[(Q_s^h - \zeta_s^h)M_s,(Q_s^h + \zeta_s^h)M_s]$, except with probability $\eps$. In the protocol below, Alice will hence abort
if $M_{{\rm left},s}$ lies outside this interval for any setting $s \in S$.

From the observed quantities $Q_s^{\rm meas}$ for different settings, Alice can obtain a lower bound on the yield of the single-photon emissions
following standard techniques used in decoy
state QKD~\cite{decoy,decoy1,decoy2,estimation}. 
Let us denote this lower bound as $\tau$. 
For honest Bob, the yield of single photons is of course just $p^{h|1}_{\clickB}$
as honest Bob always reports a round as missing if he did not observe a click. For dishonest Bob, placing a bound on this yield corresponds to
placing a bound on $1-r^{(1)}$, which in the limit of large $M$ can be seen as the probability that dishonest Bob does not choose to report
a round as missing.
Hence, we can use decoy states to obtain an estimate for the parameter $r^{(1)}$ as
\begin{align}\label{eq:newBoundR1}
r^{(1)}\leq{}1-\tau\ ,
\end{align}
even if Bob is \emph{dishonest}.
In Section~\ref{sec:concreteSource} we provide an explicit expression for $\tau$ for the case of a source emitting 
phase-randomized coherent states. 

\begin{protocol}{2}{Weak String Erasure with Errors (WSEE) using decoy states}
{Outputs: $x^m \in \sbin^m$ to Alice, 
$(\cI,z^{|\cI|}) \in 2^{[m]} \times \sbin^{|\cI|}$ 
to Bob.}\label{proto:wsedecoy}
\item[1.] {\bf Alice:} Chooses a string $x^{\hat{M}} \in_R \01^{\hat{M}}$ and basis-specifying 
string $\theta^{\hat{M}} \in_R \01^{\hat{M}}$ uniformly at random. 
\item[2.] {\bf Bob: } Chooses a basis string $\tilde{\theta}^{\hat{M}} \in_R \01^{\hat{M}}$ uniformly at random. He initializes 
$\cM \leftarrow \emptyset$.
\item[3.] In time slot $i=1,\ldots,\hat{M}$: 
\begin{enumerate}
\item {\bf Alice:}
Chooses a source setting $s_i \in S$ with probability $P_S(s_i)$.
Encodes bit $x_i$ in the basis given by $\theta_i$ (i.e., as $H^{\theta_i}\ket{x_i}$), and sends
the resulting state to Bob.
\item {\bf Bob:} Measures in the basis
given by $\tilde{\theta}_i$ to obtain outcome $\tilde{x}_i$.
If Bob obtains no click in this time slot, he records round $i$ as missing by letting $\cM \leftarrow \cM \cup \{i\}$.
\end{enumerate}

\item[4.] {\bf Bob: } Reports to Alice which rounds were missing by sending $\cM$. 
\item [4'.] {\bf Alice: } For each possible source setting $s \in S$, Alice computes the set of missing rounds 
$\cM_s = \{i \in \cM \mid s_i = s\}$. Let $\hat{M}_s = |\{j \in [\hat{M}]\mid s_j = s\}|$ be the number of rounds sent using setting $s$.
\item[5.] {\bf Alice: } For each source setting $s \in \cS$: if the number of rounds that Bob reported missing does not lie in the 
interval $[(p^h_\lostBs - \zeta^h_\lostBs)\hat{M}_s,(p^h_\lostBs + \zeta^h_\lostBs)\hat{M}_s]$, then Alice aborts 
the protocol. Otherwise, she deletes all bits from $x^{\hat{M}}$ that Bob reported missing, and all bits
that correspond to decoy state settings $s \in S$. Let $x^m \in \01^m$
denote the remaining bit string, and let $\theta^m$ be the basis-specifying string for the remaining rounds.
Let $\tilde{\theta}^m$, and $\tilde{x}^m$ be the corresponding strings for Bob.

\item[Both parties wait time $\Delta t$.]

\item[6.] {\bf Alice: } Informs Bob which rounds remain and sends the
  basis information $\theta^m$ to Bob, and outputs $x^m$.
\item[7.] {\bf Bob: } Computes $\cI \assign \{i \in [m] \mid \theta_i = \tilde{\theta}_i\}$, and outputs $(\cI,z^{|\cI|}):=(\cI,\tilde{x}
_{\cI})$.
\end{protocol}

We now state the security parameters for this protocol for the case of large $M_s = \hat{M}_s$ for each possible source. The only difference to the previous statement is that
we replace the bound on the rate~\eqref{eq:boundRate} with the bound obtained by bounding $r^{(1)}$ as in~\eqref{eq:newBoundR1}.
The parameter $M$ refers to the number of valid pulses coming from the signal setting. The decoy pulses are merely used as an estimate, 
and play no further role in the protocol. However, the probability
$\eps$ to make a correctness or security error is increased by $\eps$ for
every interval check Alice does. As she does one check per source
setting, we get a factor of $1+|S|$ increase in the error probability.

\begin{theorem}[WSEE with decoy states]\label{thm:WSEEmaintextdecoy}
Let $M = \hat{M}_{\rm signal}$.
When Bob's storage is given by $\cF=\cN^{\otimes \nu M_{\rm store}}$ for a storage 
rate~$\nu>0$, with $\cN$ satisfying the strong converse property~\cite{rs:converse}, and
having capacity~$C_\cN$ bounded by
\begin{align}
C_\cN\cdot\nu< \left(\frac{1}{2} - \delta\right)\frac{\tau}{p^{h|1}_{\clickB}}\ , 
\end{align} 
with $\tau\leq 1 - r^{(1)}$, then Protocol~1 is an $(m,\lambda(\delta),\eps(\delta),p^h_{\errB})$-weak string erasure protocol with errors 
with the following parameters:
Let $\delta\in ]0,\frac{1}{2}-C_\cN\cdot \nu[$.
Then the min-entropy rate $\lambda(\delta)$ is given by 
\begin{align}
\lambda(\delta) &=
\min_{\{r^{(n)}\}_n}
\frac{1}{m} 
\left[
\nu \cdot \gamma^\cN\left(\frac{R}{\nu}\right) M_{\rm store} - 
\sum_{n=2}^{\infty} M^{(n)}_{\rm left} \log\left(1 - p^{d,n}_{\errB}\right)\right]\ ,
\end{align}
where $\gamma^\cN$ is the strong converse
parameter of $\cN$ (see~\eqref{eq:converseParameter}) and 
the minimization is taken over all $\{r^{(n)}\}_n$ with $1-r^{(1)} \geq \tau$ such that 
$\sum_{n=1}^{\infty} r^{(n)} M^{(n)} \leq M^d_{\rm report}$ and
\begin{align}
m &= \sum_{n=1}^{\infty} M^{(n)}_{\rm left}\ , & M_{\rm store} &= p^{1}_{\rm sent} \cdot p^{h|1}_{\clickB} \cdot M\\
R &= \left(\frac{1}{2} - \delta\right)
\frac{1-r^{(1)}}{p^{h|1}_{\clickB}}\ , & M^d_{\rm report} &= (p^h_\lostB - p^d_\lostB)M\ ,
\end{align}
for sufficiently large $M$.
The error has the form
\begin{align}
\eps(\delta) \leq (1+|S|) \cdot 2\exp\left( 
- \frac{\delta^2}{512(4 + \log\frac{1}{\delta})^2} \cdot 
\tau\ p^1_\sent M
\right)\ .
\end{align}
\end{theorem}

\section{Oblivious transfer from WSEE}\label{sec:otFromWSEE}

\begin{figure}
\begin{center}
\includegraphics{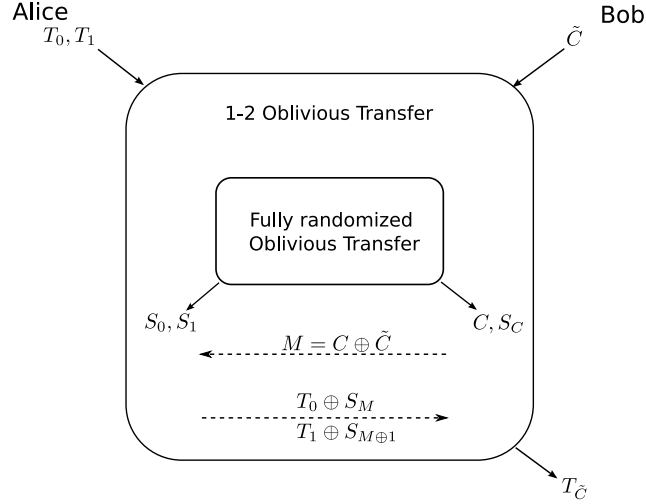}
\caption{1-2 oblivious transfer from fully randomized transfer by sending additional messages given by the dashed lines.}
\label{fig:FROTtoOT}
\end{center}
\end{figure}

We now show how to obtain oblivious transfer from WSEE.  Here we implement a fully randomized oblivious transfer protocol
(FROT), which can easily be converted into 1-2 oblivious transfer as shown in Figure~\ref{fig:FROTtoOT}.
We now give an informal description of this task, and refer to~\cite{noisy:new} for a
formal definition.

\begin{definition}[Informal]
An \emph{$(\ell,\eps)$-fully randomized oblivious transfer protocol (FROT)} is a protocol between two parties, Alice and Bob, satisfying the following properties:
\begin{description}
\item[Correctness:] If both parties are honest, then Alice obtains two random strings $S_0, S_1 \in \01^{\ell}$, and Bob
obtains a random choice bit $C \in \01$ as well as $S_C$.
\item[Security for Alice:] If Alice is honest, then there exists $C
  \in \01$ such that given $S_C$, Bob cannot learn anything about
$S_{1-C}$, except with probability $\eps$.
\item[Security for Bob:] If Bob is honest, then Alice learns nothing about $C$.
\end{description}
\end{definition}

\subsection{Ingredients} 

\subsubsection{Suitable error-correcting codes}
To deal with the bit-flip errors in the weak string erasure we need to
augment the protocol of~\cite{noisy:new} with an additional
error-correction step as in~\cite{noisy:robust}. That is, Alice has to
send some small amount of error-correcting information to Bob. The
challenge we face is to ensure that security is preserved: Recall that
if Bob is dishonest, we assume a worst-case scenario where he does not
experience any transmission errors and he can perform perfect quantum
operations. Hence, he could use this additional error-correcting
information to correct some of the errors caused by his noisy quantum
storage. On the other hand, if Alice is dishonest, we have to
guarantee that the error-correcting process does not allow her to gain
any information about the choice bit $C$. This last requirement can be
achieved by using a \emph{one-way} (or \emph{forward})
error-correcting code in which only Alice sends information to Bob
\footnote{This is in contrast to QKD where common solutions typically
  use \emph{interactive} error correction, such as the cascade
  scheme~\cite{BS93}.}. Let $\{C_n\}$ be a family of linear error-correcting codes
of length $n$ capable of efficiently correcting $p_{\rm err} \cdot n$
errors. For a $k$-bit string $x^k$, error correction is done by
sending the syndrome information $\syn(x^k)$ to Bob who can then
efficiently recover $x^k$ from his noisy string $\cE_{p_{\rm
    err}}(x^k)$. For instance, low-density parity-check (LDPC) codes
can correct a $k$-bit string, where each bit flipped with probability
$p_{\rm err}$, by sending at most $1.2\cdot h(p_{\rm err})\cdot k $
bits of error-correcting information~\cite{ELAB09}.

\subsubsection{Interactive hashing}
\begin{figure}[h]
\begin{center}
\includegraphics{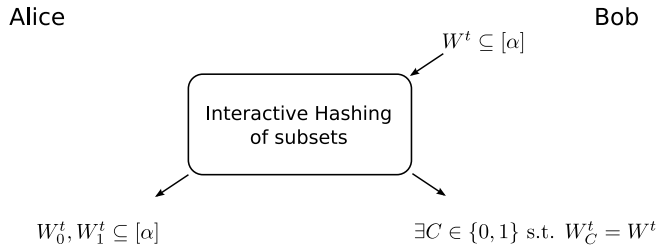}
\end{center}
\caption{Interactive hashing} \label{fig:IH}
\end{figure}

Apart from an error-correcting code, the protocol below requires three
classical ingredients that need to be implemented: First, we need to
use the primitive of interactive hashing of subsets. This is a
classical protocol in which Bob holds as input a subset $W^t \subseteq
[\alpha]$ (where $\alpha$ is some natural number) and Alice has no
input. Both Alice and Bob receive two subsets $W^t_0, W^t_1 \subseteq
[\alpha]$ as outputs, where there exists some $C \in \01$ such that
$W^t_C = W^t$ as depicted in Figure~\ref{fig:IH}.
Informally, security means that Alice does not learn $C$, and
$W^t_{1-C}$ is chosen almost at random from the set of all possible
subsets of $[\alpha]$. That is, Bob has very little control over the
choice of $W^t_{1-C}$. Here we restrict ourselves to this definition
and refer to~\cite{noisy:new} for a formal definition. In order to
perform interactive hashing, we describe below how to encode the input
subsets into a $t$-bit string. Intuitively, interactive hashing can be
done by Alice asking Bob for random parities of his $t$-bit string
$W$. After $t-1$ linearly independent queries, there are only two
possible strings left: one of which is Bob's original input, the other
one is pretty much out of his control. A concrete protocol for
interactive hashing can be found, for instance, in~\cite{ding}.

\subsubsection{Encoding of subsets}
The second ingredient we need is thus an encoding of subsets as bit
strings.  More precisely, we map $t$-bit strings to subsets
using~$\Enc:\sbin^t\rightarrow \cT$, where~$\cT$ is the set of all
subsets of $[\alpha]$ of size~$\alpha/4$. Here we assume without loss of
generality that $\alpha$~is a multiple of~$4$.  The encoding $\Enc$ is
injective, that is, no two strings are mapped to the same subset.
Below, we furthermore choose~$t$ such that $2^t\leq
\binom{\alpha}{\alpha/4}\leq 2\cdot 2^t$.  This means that
not all possible subsets are encoded, but at least half of them. We
refer to~\cite{ding,savvides:diss} for details on how to obtain such
an encoding.  

\subsubsection{Two-universal hashing}
Finally, we require the use of two-universal hash
functions for privacy amplification as they are also used in QKD~\cite{renato:diss}.  
Any implementation used for QKD may be used here.
Below, we use $\cR$ to denote the set of possible hash functions, and use 
$\Ext(X,R)$ to represent the output of the hash function given by $R$ when applied to the
string $X$.

\subsection{Protocol}
\begin{figure}
\begin{center}
\includegraphics{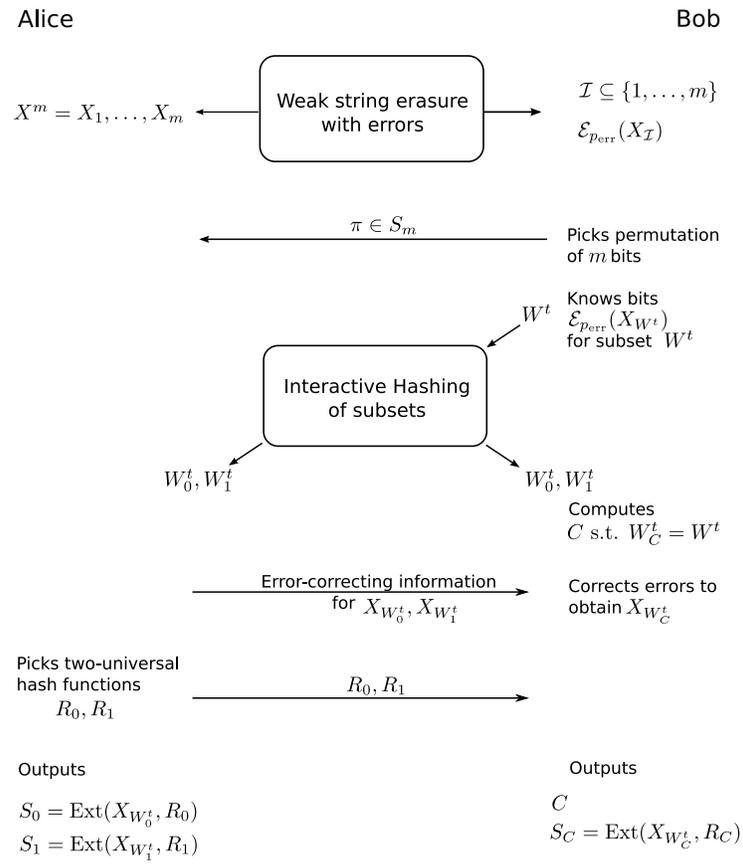}
\end{center}
\caption{Conceptual steps in the protocol for FROT from WSEE.}
\label{fig:OToutline}
\end{figure}

Before providing a detailed
description of the protocol, we first give a description of
the different steps involved in Figure~\ref{fig:OToutline}. 

\begin{protocol}{3}{WSEE-to-FROT}{
Parameters: Integers $m,\beta$ such that  $\alpha := m/\beta$ 
is a multiple of~$4$. Set $t:=\alpha/2$. 
Outputs: $(s_0^\ell,s_1^\ell) \in
\{0,1\}^{\ell} \times \01^\ell$ to Alice, and $(c,y^\ell) \in \{0,1\} \times \{0,1\}^\ell$
to Bob }
\item {\bf Alice and Bob: }  Execute $(m,\lambda,\eps,p_{\rm err})$-WSEE.
Alice obtains a string $x^m \in \{0,1\}^m$, Bob a set $\cI \subset [m]$
and a string $s = \cE_{p_{\rm err}}(x_\cI)$. If $|\cI| < m/4$, Bob aborts.
Otherwise, he
randomly truncates $\cI$ to the size $m/4$, and deletes the
corresponding values in $s$.

We arrange $x^m$ into a matrix ${\bf z} \in \Mat_{\alpha \times \beta}(\sbin)$, by 
${\bf z}_{j,k} := x_{(j-1) \cdot \beta + k}$ for
$(j,k) \in [\alpha] \times [\beta]$.
 
\item {\bf Bob: }
\begin{enumerate}
\item Randomly chooses a string $w^t\in_R \{0,1\}^t$ corresponding to an encoding  of a subset $\Enc(w^t)$ of $[\alpha]$ with 
$\alpha/4$ elements.
\item Randomly partitions the $m$~bits of $x^m$ into $\alpha$ blocks
  of $\beta$~bits each: He randomly chooses a permutation
  $\pi:[\alpha]\times[\beta]\rightarrow [\alpha]\times [\beta]$ of the
  entries of ${\bf z}$ such that he knows $\pi({\bf z})_{\Enc(w^t)}$
  (that is, these bits are permutation of the bits of~$s$). Formally,
  $\pi$ is uniform over permutations satisfying the following
  condition: for all $(j, k) \in [\alpha] \times [\beta]$ and $(j',
  k') := \pi (j,k)$, we have $(j-1) \cdot \beta + k \in \cI$ if and only if $j' \in \Enc(w^t)$.
  
\item Bob sends $\pi$ to Alice.
\end{enumerate}
\item {\bf Alice and Bob:} Execute interactive hashing with Bob's input
equal to $w^t$. They obtain $w_0^t,w_1^t \in \{0,1\}^t$ with
$w^t\in\{w_0^t,w_1^t\}$. 
\item {\bf Alice: } Sends error-correcting information for every block
  in $\Enc(w_0^t)$ and $\Enc(w_1^t)$, i.e., $\forall j \in \Enc(w_0^t)
  \cup \Enc(w_1^t)$, Alice sends $\Syn(\pi({\bf z})_j)$ to Bob.
\item {\bf Alice: } Chooses $r_{0}, r_{1} \in_R \cR$ and sends them to Bob.
\item {\bf Alice: } Outputs
$(s_0^\ell,s_1^\ell) \assign (\Ext(\pi({\bf z})_{\Enc(w_0^t)},r_0),
\Ext(\pi({\bf z})_{\Enc(w_1^t)},r_1))$.
\item {\bf Bob: } 
Computes $c$, where $w^t=w_c^t$, and $\pi({\bf
z})_{\Enc(w^t)}$ from $s$. 
Performs error correction on the blocks of $\pi({\bf z})_{\Enc(w^t)}$.
He outputs $(c,y^\ell) \assign
(c,\Ext(\pi({\bf z})_{\Enc(w^t)},r_c))$.
\end{protocol}

When using WSEE to obtain FROT, Protocol 3 achieves the following
parameters. The proof of this statement
can be found in Appendix~\ref{app:OT}. 

\begin{theorem}[Oblivious transfer]
For any constant $\omega \geq 2$ and $\beta \geq \max\{67,256 \omega^2/\lambda^2\}$, the protocol WSEE-to-FROT 
implements an $(\ell,41 \cdot 2^{-\frac{\lambda^2}{512 \omega^2 \beta} m} + 2 \eps)$-FROT from one 
instance of $(m, \lambda, \eps,p_{\rm err})$-WSEE, where
$$\ell := \left \lfloor \left(\left(\frac{\omega-1}{\omega}\right)\frac{\lambda}{8}-\frac{\lambda^2}{512 \omega^2\beta} - 
\frac{1.2 \cdot h(p_{\rm err})}{8}\right)m - \frac{1}{2} \right \rfloor$$.
\end{theorem}

\begin{figure}[h]
\begin{center}
\includegraphics[scale=0.5]{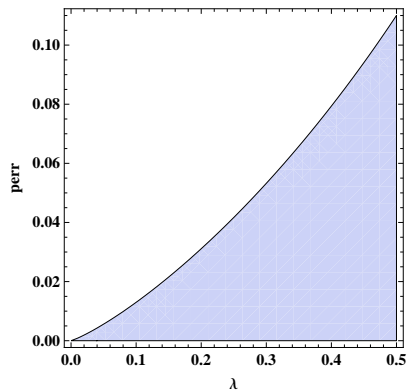}
\end{center}
\caption{(Color online) Security can be achieved if $(p_{\rm err},\lambda)$ lies in the shaded region, where we chose a very large value of
$\omega = 100000$.}
\label{fig:lambdaVSperr}
\end{figure}

The parameter $\omega$ appearing in the theorem above is an additional 
parameter that we can tune to trade off a higher rate of OT, against an error that decays more slowly.
Our choice of $\omega$ will thereby depend on the error $p_{\rm err}$: Note that for large values of $\omega$, 
we can essentially achieve security as long as $\lambda > h(p_{\rm err})$ (see Figure~\ref{fig:lambdaVSperr}). Of course, this requires us to use many
more rounds to be able to achieve the desired block size $\beta$, as well as to make the error sufficiently small again.
Using more rounds, however, may be much easier than to decrease the bit error rate of the channel.

\section{Security for two concrete implementations}\label{sec:concreteSource}

We now show how our security analysis applies to two particular experimental setups using weak coherent pulses or a parametric down conversion source.
Unlike in QKD, our protocols are particularly interesting at short distance, where one may use visible light for which better detectors exist.

\subsection{Phase-randomized weak coherent pulses}\label{sec:wcp}

\subsubsection{Experimental setup and loss model}
\label{sec_wcp_pract}
\begin{figure}
\begin{center}
\includegraphics[scale=0.8]{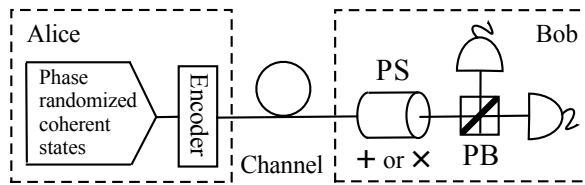}
\caption{Experimental setup with phase-randomized weak coherent pulses. The Encoder 
codifies the BB84 signal information. The polarization shifter (PS) allows to change the polarization basis 
(computational basis $+$ or Hadamard basis $\times$) of the measurement as desired. The polarization analyzer consists of a polarizing beam splitter 
(PB) and two threshold detectors. The PB discriminates the two orthogonal polarized modes.
\label{figure:wcp}}
\end{center}
\end{figure}

We first consider a phase-randomized weak coherent source. The 
basic setup for Alice and Bob is illustrated in Figure~\ref{figure:wcp}.
The signal states sent by Alice can be described as
\begin{equation}\label{mv}
\rho_k=e^{-\mu} \sum_{n=0}^{\infty} \frac{\mu^n}{n!} \outp{n_k}{n_k},
\end{equation}
where the signals $\ket{n_k}$ denote Fock states with $n$ photons in one of the four possible polarization states 
of the BB84 scheme, which are labeled with the index $k$. 

On the receiving side, we shall assume that honest Bob uses an
active-basis-choice measurement setup. It consists of a polarization
analyzer and a polarization shifter which effectively changes the
polarization basis of the subsequent measurement. The polarization
analyzer has two threshold detectors, each monitoring the output of a
polarizing beam splitter. These detectors are characterized by their
\emph{detection efficiency} $\eta$ and their \emph{dark-count
  probability} $\pdark$. Notice that we include all sources of loss in
the system (including channel loss, coupling loss in Alice's and Bob's
laboratory etc.) in the definition of the detection efficiency
$\eta$~\footnote{In this work, we are \emph{not} considering the
  detection efficiency mismatch problem and detector-related attacks
  such as time-shift attacks~\cite{QFLM07,Zhao08} or faked-state
  attacks~\cite{MAS06,MS08}.  We remark that security proofs for QKD
  schemes that take into account such detection efficiency mismatch do
  exist, see e.g.~\cite{FTQLM09}.}. 
For the case of honest Alice and Bob,
the {\it overall} transmittance, $\eta$, is a product.
i.e.,  $\eta = \eta_A \eta_{channel } \eta_B \eta_D $
where $\eta_A$ is the transmittance on Alice's side,
$\eta_{channel}$ is the channel transmittance,
$\eta_B$ is the transmittance on Bob's side (excluding detection inefficiency)
and $\eta_D$ is the detector efficiency defined previously in the introductory section.
Recall, from the introductory section, that $\eta_D$ is about $10\%$
for telecom wavelengths and $ 70\%$ for visible wavelengths.

Now, for some practical set-ups (such as short-distance
free-space with visible wavelength), it is probably technologically feasible to
achieve $\eta_A \eta_{channel} \eta_B$ of order 1, say 50 percents.
In more detail, in some set-ups (e.g. with a weak coherent state source),
Alice may compensate for her internal loss by characterizing it
and then simply turning up the intensity of her laser. In those cases,
she may effectively set $\eta_A =1$.
Now, for short-distance applications, $\eta_{channel}$, can be made
of order $1$. All that is required to achieve $\eta_A \eta_{channel} \eta_B$ of order 1
is to reduce Bob's internal loss, thus boosting $\eta_B$ to order 1.
For simplicity, we consider that both detectors have equal parameters. Since we absorb all terms into the detector
inefficiency, we simply refer to this as $\eta$. 

As in QKD~\cite{lopreskill:randomize2}, the fact that each signal
state is phase-randomized is an important element for our security
analysis. It allows us to argue that, without loss of generality, a
dishonest Bob always performs a {\it quantum nondemolition} (QND)
measurement of the total number of photons contained in each pulse
sent by Alice.  Hence, we can analyze the single-photon pulses
separately from the multi-photon pulses, which makes an important
difference for Bob's cheating capabilities.  In Appendix C, we compute
all relevant probabilities to evaluate security in this
scenario. These probabilities are summarized in the following
Table~\ref{tab:parametersWCP}. For completeness, we explicitly state
some parameters which we need in order to evaluate the error
probability $p^h_{\errB}$. These parameters are: the probability that
Bob makes an error due to dark counts alone ($p_\derr$), the signal
alone ($p_{\serr}$), and the probability that he makes an error due to
dark counts and the signal ($p_{\dserr}$), as well as the probability
that a signal alone produces no click in Bob's side ($p^h_{\slostB}$).

\begin{table}[h]
\begin{center}
\begin{tabular}{|l|l|}
\hline
Parameter & Value\\
\hline
$p^1_\src$ & $e^{-\mu} \mu$\\
\hline
$p^1_\sent$ & $p^1_\src$\\
\hline
$p^{h|1}_{\clickB}$ & $\eta + (1-\eta) \pdark (2 - \pdark)$\\
\hline
$p^{d,n}_{\rm B, err}$ & $0$\\
\hline
$p^{d}_\lostB$ & $e^{-\mu}$\\
\hline
$p^{h}_{\slostB}$ & $e^{-\mu} \sum_{n=0}^{\infty} \frac{\mu^n}{n!} (1-\eta)^n=e^{-\eta\mu}$\\
\hline
$p^{h}_{\lostB}$ & $p^h_{\slostB} - p^h_{\slostB} \pdark(2-\pdark)$\\
\hline
$p_\derr$ & $\pdark(1-\pdark) + \pdark^2/2$\\
\hline
$p_\dserr$ & $(1-p^h_\slostB) 
\left((1-\edet)\frac{\pdark}{2} + \edet\cdot \pdark (\frac{3}{2} -\pdark)\right)$\\
\hline
$p^h_\serr$ & $\edet (1 - p^h_{\slostB})$\\
\hline
$p^h_{\errB}$ & $p^h_\serr (1 - \pdark (2-\pdark))+ p^h_{\slostB} 
p_\derr + p_\dserr$\\
\hline
\end{tabular}
\end{center}
\caption{Summary of the probabilities for phase-randomized weak
  coherent pulses \label{tab:parametersWCP}}
\end{table}

\subsubsection{Security parameters}\label{sec:basisParameters}

To evaluate the probabilities above we assume that 
$
p_{\dark} = 0.85 * 10^{-6}\ ,
$
and use 
$e_{\rm det} = 0.033$ as a very conservative number on a distance of 122 km~\cite{GYS04}.

\paragraph{Weak string erasure}\label{sec:wcpWSE1}

\begin{figure}[h]
\begin{minipage}[t]{0.45\textwidth}
\begin{center}
\includegraphics[width=5cm]{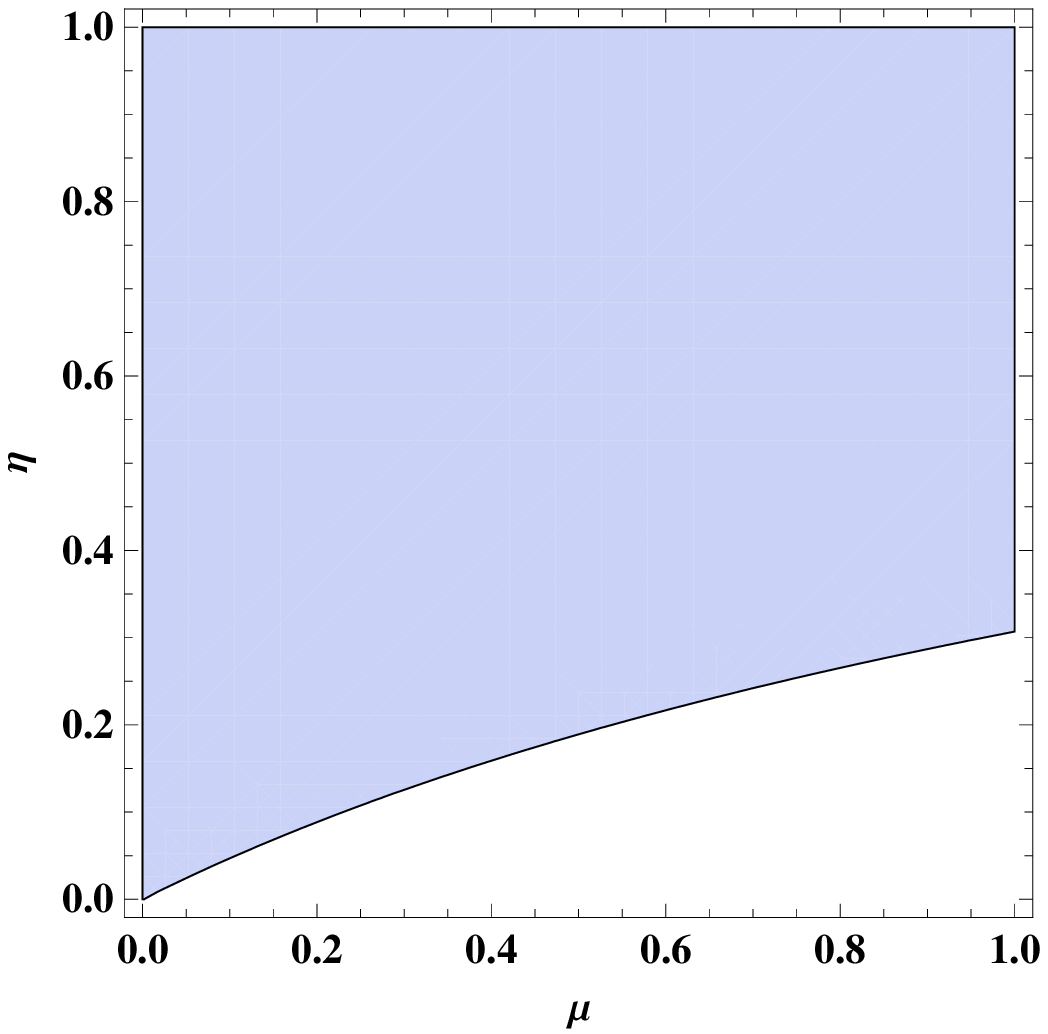}
\caption{(Color online) Security possible for $(\eta,\mu)$ in the shaded region where
\eqref{eq:cond1} is fulfilled. Our proof does not apply to parameters
in the region below the curve. For the shaded region above the curve,
additional conditions such as \eqref{eq:cond2} are checked in the
following Figure~\ref{fig:wcpCond2numu}.}
\label{fig:wcpCond1numu}
\end{center}
\end{minipage}\hspace{5mm}
\begin{minipage}[t]{0.45\textwidth}
\begin{center}
\includegraphics[width=5cm]{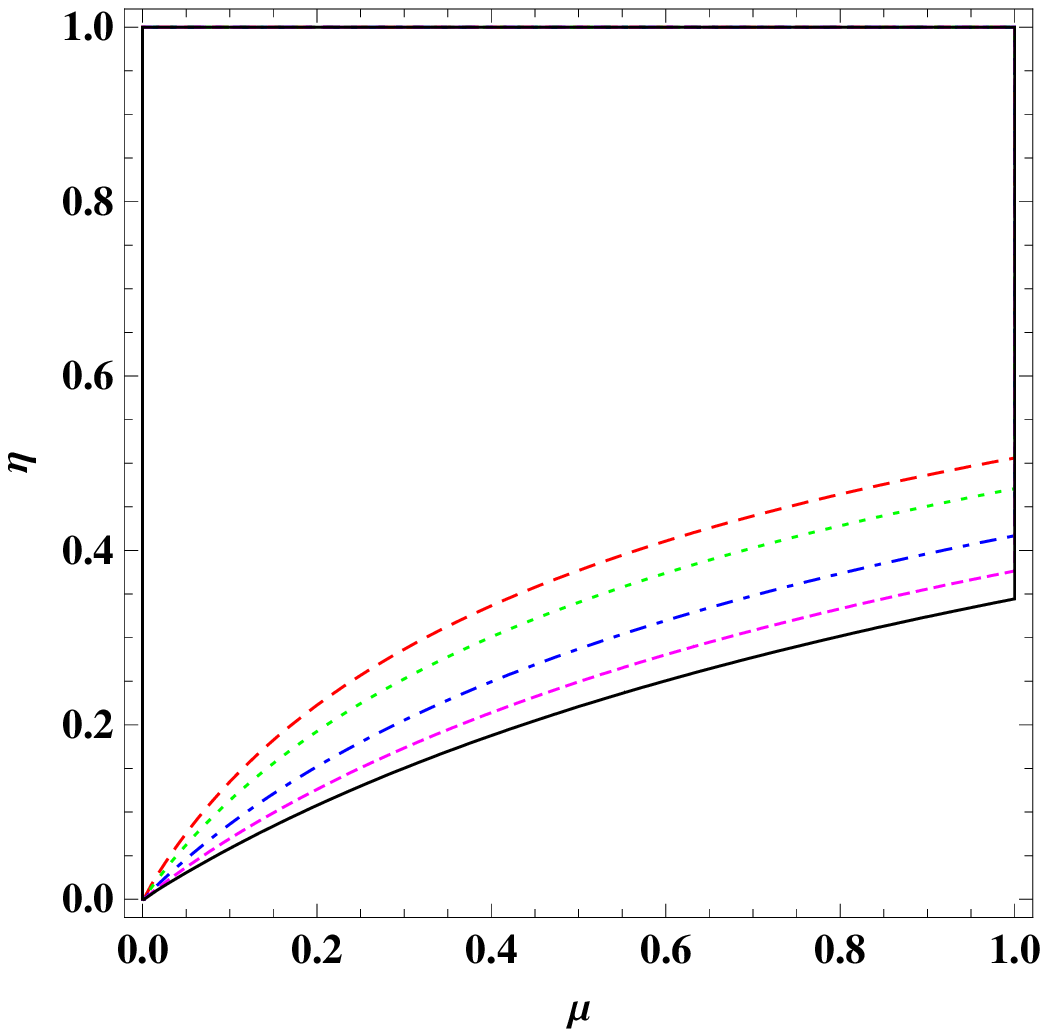}
\caption{(Color online) Security possible for $(\eta,\mu)$ in the upper enclosed regions for a low storage noise of $r = 0.9$ and 
storage rates $\nu$ of $1/2$ (dashed red line), $0.45$ (dotted green line), $0.35$ (dot dashed blue line), $0.25$ (large dashed magenta line),
$0.15$ (solid black line) (satisfying~\eqref{eq:cond1} and~\eqref{eq:cond2}).}
\label{fig:wcpCond2numu}
\end{center}
\end{minipage}
\end{figure}

We now investigate the security of $(m,\lambda,\eps,p_{\rm err})$-weak
string erasure, when using a weak coherent source.  Before examining
the weak string erasure rate $\lambda$ that one can obtain for some
set of source parameters, we first consider when security can be
obtained in principle (i.e., when~\eqref{eq:cond1}
and~\eqref{eq:cond2} are satisfied) as a function of the mean photon
number $\mu$, the detection efficiency $\eta$, the storage rate $\nu$
and the amount of storage noise. Our examples here focus on the
depolarizing channel with parameter $r$ as defined
in~\eqref{eq:depol}.  First of all, Figure~\ref{fig:wcpCond1numu}
tells us when security is possible at all, independently of the amount
of storage noise.  We then examine a particular example of storage
noise and storage rates in Figure~\ref{fig:wcpCond2numu}.  This shows
that even for low storage noise, we can hope to achieve security for
many source settings. Note that this plot is merely an example, and of
course does not rule out security of other forms of storage noise or
other storage rates. The following plots have been made using
Mathematica, and the corresponding files used are available upon
request.

\begin{figure}[h]
\begin{center}
\includegraphics[width=8cm]{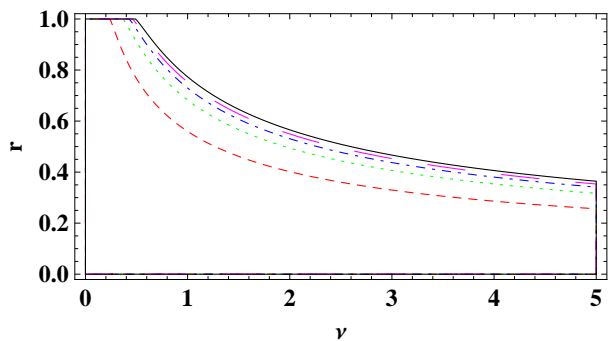}
\caption{(Color online) Security for $(r,\nu)$ below the lines for $\mu = 0.3$ and
detection efficiencies $\eta$: $0.7$ (solid black line), $0.5$ (large dashed magenta line), $0.4$ (dot dashed blue line), 
$0.3$ (dotted green line), $0.2$ (dashed red line).}
\label{fig:wcpNuvsR}
\end{center}
\end{figure}

We now consider when conditions~\eqref{eq:cond1} and~\eqref{eq:cond2} can be satisfied in terms of the 
amount of noise in storage given by $r$, and the storage rate $\nu$ for some typical parameters in an experimental setup.
Figure~\ref{fig:wcpNuvsR} shows us, that there is a clear trade-off between $r$ and $\nu$ 
dictating when weak string erasure can be obtained from our analysis, but typical parameters of the source
move us well within a possible region. 

\begin{figure}[h]
\begin{center}
\includegraphics[width=7cm]{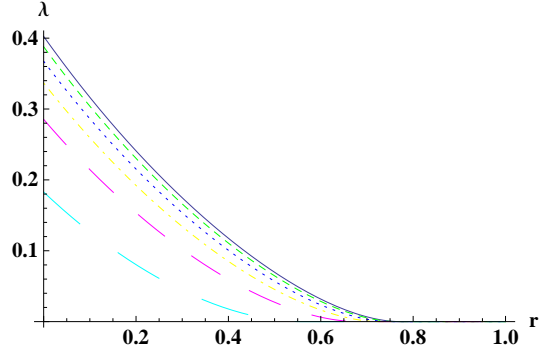}
\caption{(Color online) The WSE rate $\lambda$ in terms of the amount of depolarizing noise $r$ where $\mu = 0.3$, and
a variety of detection efficiencies $\eta$: $0.7$ (solid black line), $0.6$ (dashed red line), $0.5$ (dotted blue line), 
$0.4$ (dot dashed yellow line), $0.3$ (large dashed magenta line) and $0.2$ (larger dashed turquoise line).}
\label{fig:wcpLambdaExample}
\end{center}
\end{figure}

Now that we have established that secure weak string erasure can be obtained for a reasonable choice of parameters, it remains to establish
the weak string erasure rate $\lambda$. This parameter cannot be read off explicitly, but is determined by the optimization
problem given in~\eqref{eq:lambdaOptimize}. To gain some intuition about the magnitude of this parameter we plot it in Figure~\ref{fig:wcpLambdaExample} 
for various choices of experimental settings, and a storage rate of $\nu=1$.
This shows that even for a very high storage rate, there is a positive rate of $\lambda$ for many reasonable settings.
Of course $\lambda$ can be larger if we were to consider a lower storage rate.

\begin{figure}[h]
\begin{minipage}[t]{0.45\textwidth}
\begin{center}
\includegraphics[width=5cm]{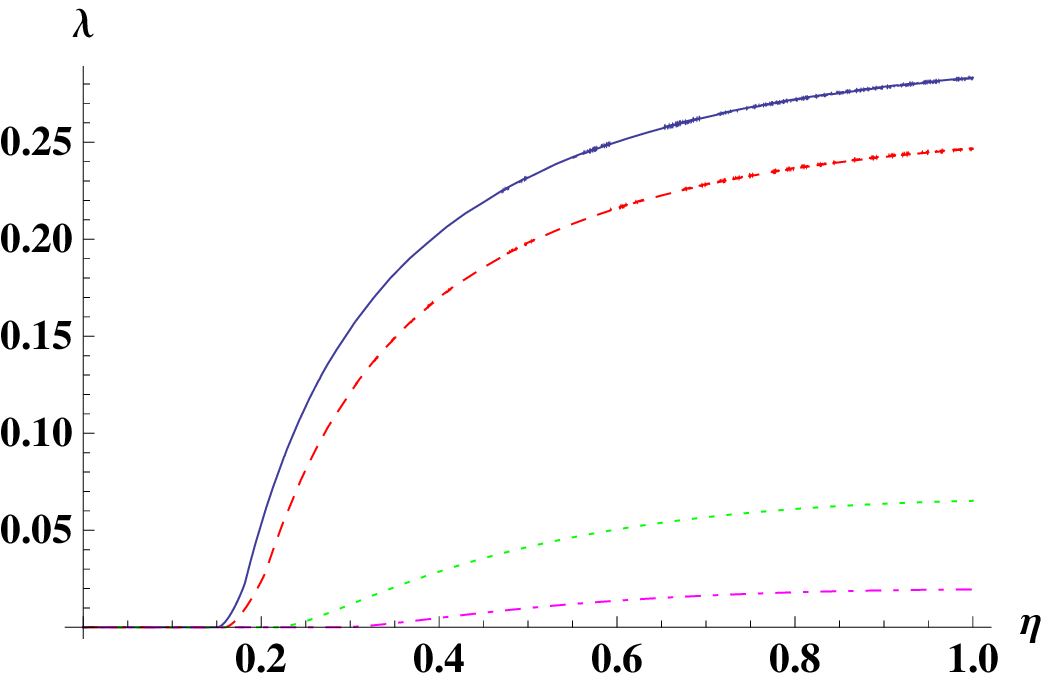}
\caption{(Color online) The WSE rate $\lambda$ in terms of the detection efficiency $\eta$ for $r=0.8$ and storage rates
$\nu$: $1/5$ (solid blue line), $1/4$ (dashed red line), $1/2$ (dotted green line), $2/3$ (dot dashed magenta line).}
\label{fig:wcpLambdavsEta}
\end{center}
\end{minipage}\hspace{5mm}
\begin{minipage}[t]{0.45\textwidth}
\begin{center}
\includegraphics[width=5cm]{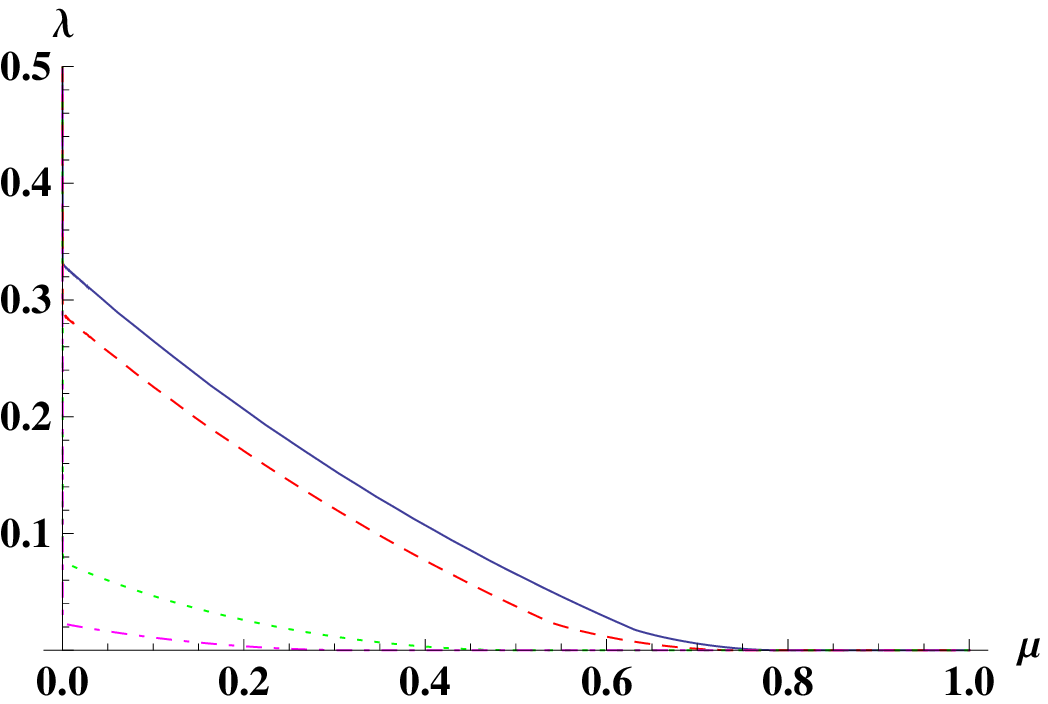}
\caption{(Color online) The WSE rate $\lambda$ in terms of the mean photon number $\mu$ for $r=0.8$ and $\nu$ as in Figure~\ref{fig:wcpLambdavsEta}.}
\label{fig:wcpLambdavsMu}
\end{center}
\end{minipage}
\end{figure}

To gain further intuition into the role that the different parameters play in determining the rate $\lambda$, 
we investigate the trade-off between $\lambda$ and the detection efficiency $\eta$ in Figure~\ref{fig:wcpLambdavsEta},
and the trade-off between $\lambda$ and the mean photon number $\mu$ in Figure~\ref{fig:wcpLambdavsMu} for
some choices of storage noise $r$ and storage rate $\nu$. 

\paragraph{1-2 oblivious transfer}\label{sec:wcpOT1}

\begin{figure}
\begin{center}
\includegraphics[width=5cm]{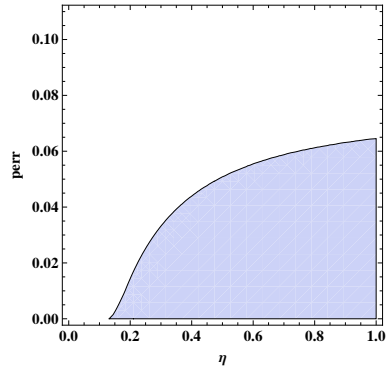}
\caption{(Color online) Security for $(p_{\rm err},\eta)$ in the shaded region for example parameters
$r=0.4$, $\nu=1/5$ and large $\omega = 100000$.}
\label{fig:wcpErrvsEta}
\end{center}
\end{figure}

We can now consider the security of $(\ell, \eps)$-oblivious transfer based on weak string erasure implemented using a weak coherent source.
The parameter which is of most concern to us here is the bit error rate $p_{\rm err} = p^h_{\errB}/(1-p^{h}_{\lostB})$.
As we already saw in Figure~\ref{fig:lambdaVSperr},
this error cannot be arbitrarily large for a fixed value of the WSE rate $\lambda$.
In a practical implementation, this translates into a trade-off between the bit error $p_{\rm err}$ and the efficiency $\eta$ as shown in Figure~\ref{fig:wcpErrvsEta},
where for now we treat $p_{\rm err}$ as an independent parameter to get an intuition for its contribution.

\begin{figure}
\begin{center}
\includegraphics[scale=0.8]{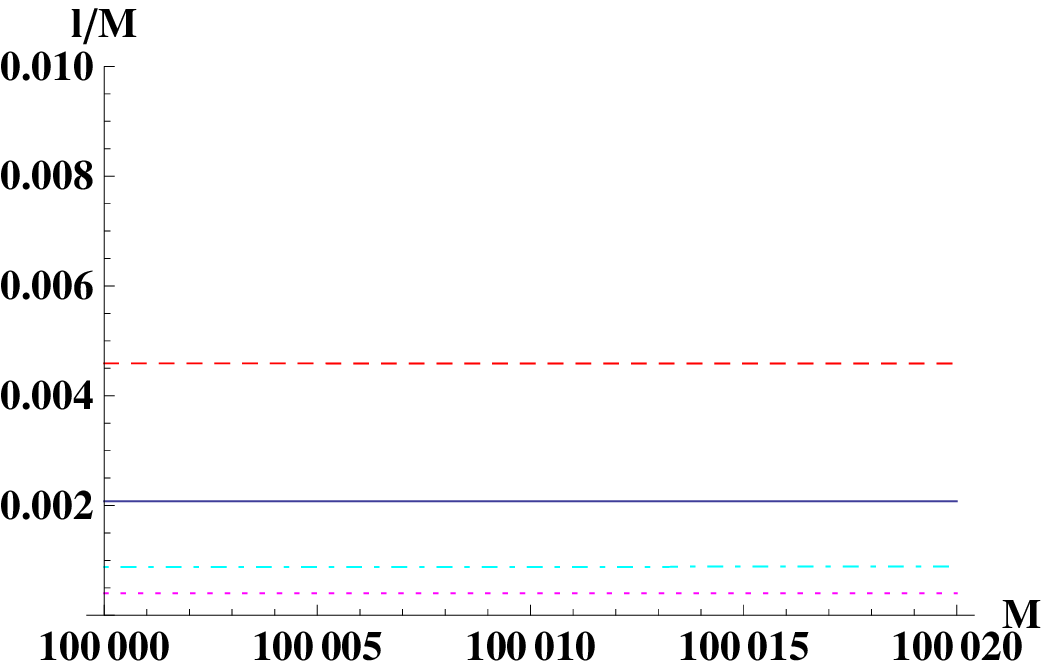}
\caption{(Color online) The rate $\ell/M$ of oblivious transfer for a large number of valid pulses $M$ for parameters $\omega=1000$ and
($\mu=0.15$, $\eta = 0.3$, $r=0.1$, $\nu=1/10$, solid blue line),
($\mu=0.4$, $\eta=0.7$, $r=0.1$, $\nu=1/10$, dashed red line),
($\mu=0.15$, $\eta=0.7$, $r=0.7$, $\nu=1/4$, dotted magenta line),
($\mu=0.2$, $\eta=0.7$, $r=0.4 $, $\nu=1/3$, light blue line).}
\label{fig:wcpRatePlot}
\end{center}
\end{figure}

Of course $p_{\rm err}$ is not an independent parameter, but depends on $\mu$, $\eta$ and most crucially on $\edet$. 
Figure~\ref{fig:wcpRatePlot} shows how many bits $\ell$ of 1-2 oblivious transfer we can hope to obtain per valid pulse $M$
for very large $M$. The parameter $\mu$ has thereby been chosen to obtain a high rate when all other parameters were fixed.
We will also refer to $\ell/M$ as the oblivious transfer rate. 
As expected, we can see that this rate does of course depend greatly on the efficiency $\eta$, 
but also on the storage noise and on the storage rate.

\subsubsection{Parameters using decoy states}\label{sec:wcpDecoy}

We now analyze the scenario where Alice sends decoy states. In
particular, let us consider a simple system with only two decoy
states: vacuum and a weak decoy state with mean photon number
$\hat{\mu}$. The mean photon number of the signal states will be denoted as
$\mu$. Moreover, we select $\hat{\mu}<\mu$.  Without loss of generality, we
hence use labels $\cS = \{{\rm vac}, \hat{\mu}, \mu\}$ for the possible
settings of the source.  Furthermore, we assume that Alice chooses one
of these settings uniformly at random, that is $P_S(s) = 1/3$ for all
$s \in \cS$. This may not be optimal, but due to the large number of
parameters we will limit ourselves to this choice.
Since $p^n_\src = p^n_\sent$ for the case of a phase-randomized weak coherent source, 
we can write for honest Bob
\begin{align}
Q^h_{\rm vac} &= p^{h|0}_\clickB\ ,\\
Q^h_{\hat{\mu}} &= e^{-\hat{\mu}} \sum_{n=0}^{\infty} \frac{\hat{\mu}^n}{n!} p^{h|n}_\clickB\ ,\\
Q^h_\mu &= e^{-\mu} \sum_{n=0}^{\infty} \frac{\mu^n}{n!} p^{h|n}_\clickB\ .\\
\end{align}
For the typical channel model, that is, if Bob were honest,
we furthermore have 
\begin{align}
p^{h|0}_\clickB &= 2 \pdark (1-\pdark) + \pdark^2\ ,\\
p^{h|n}_\clickB &= 1 - (1-p^{h|0}_\clickB)(1-\eta)^n\ .\\
\end{align}
For simplicity, when calculating the value of the parameter
$p^{h|0}_\clickB$ we have only considered the noise arising from dark
counts in the detectors. In a practical situation, however, there
might be other effects like stray light that also contribute to the
final value of $p^{h|0}_\clickB$. Still, from her knowledge of the
experimental setup, Alice can always make a reasonable estimate of the
maximum tolerable value of $p^{h|0}_\clickB$ such that the protocol is
not aborted and the analysis is completely analogous.
Furthermore, we have assumed that the losses come mainly from the finite detection efficiency of the detectors, since 
the communication distance will be typically quite short. 

To estimate a lower bound on the yield of single photons we follow the
procedure proposed in~\cite{estimation}.  Note, however, that many
other estimation techniques are also available, like, for instance,
linear programming tools~\cite{linear}. In the asymptotic case we
obtain~\cite{estimation}
\begin{align}\label{eq:largeNYBound}
(1-r^{(1)}) \geq \hat{\tau} \qquad \mbox{ with } \hat{\tau} := 
\frac{\mu}{\mu \hat{\mu} - \hat{\mu}^2} \left(Q^h_{\hat{\mu}} e^{\hat{\mu}} - Q^h_\mu e^{\mu} \frac{\hat{\mu}^2}{\mu^2} - \frac{\mu^2 - \hat{\mu}^2}{\mu^2} Q^h_{\rm vac}\right)\ ,
\end{align}
where we used the fact that for honest Bob $p^{h|1}_{\clickB} = 1-r^{(1)}$ in the limit
of large $M$, as Bob will decide to report any round as missing that he did not receive.
In Protocol 2 we have that
conditioned on the event that Alice does not abort the protocol
\begin{align}
Q_{\rm vac}^{\rm meas} &\in [(Q^h_{\rm vac}- \zeta_0),(Q^h_{\rm vac} + \zeta_0)]\ ,\\
Q_{\hat{\mu}}^{\rm meas} &\in [(Q^h_{\hat{\mu}} - \zeta_{\hat{\mu}}),(Q^h_{\hat{\mu}} + \zeta_{\hat{\mu}})]\ ,\\
Q_\mu^{\rm meas} &\in [(Q^h_\mu - \zeta_\mu),(Q^h_\mu + \zeta_\mu)]\ ,
\end{align}
where $\zeta_0 = \sqrt{\ln(2/\eps)/(2M_0)}$, $\zeta_{\hat{\mu}} = \sqrt{\ln(2/\eps)/(2M_{\hat{\mu}})}$ , and $\zeta_\mu = \sqrt{\ln(2/\eps)/(2 M_\mu)}$.
We can hence bound
\begin{align}\label{eq:finiteTauBound}
(1-r^{(1)}) \geq \tau \qquad \mbox{ with } \tau := 
\frac{\mu}{\mu \hat{\mu} - \hat{\mu}^2} \left((Q^h_{\hat{\mu}} - 2 \zeta_{\hat{\mu}}) e^{\hat{\mu}} - (Q^h_\mu + 2\zeta_\mu) e^{\mu} \frac{\hat{\mu}^2}{\mu^2} - \frac{\mu^2 - \hat{\mu}^2}{\mu^2} (Q^h_{\rm vac} +2 \zeta_0)\right)\ ,
\end{align}
which in the limit of large $M_0$, $M_\mu$ and $M_{\hat{\mu}}$ gives us~\eqref{eq:largeNYBound}.
The factor $2$ in the equation~\eqref{eq:finiteTauBound} 
above stems from the fact that Alice still accepts a value at the
upper (or lower) edge of the interval such as $Q^h_{\rm \mu} +
\zeta_\mu$. In this case however, the real parameter $\hat{Q}^h_{\rm
\mu}$ is possibly as high as $Q^h_{\rm \mu} +
2 \zeta_\mu$. 

\subsubsection{Weak string erasure}

\begin{figure}[h]
\begin{minipage}[t]{0.45\textwidth}
\begin{center}
\includegraphics[width=5cm]{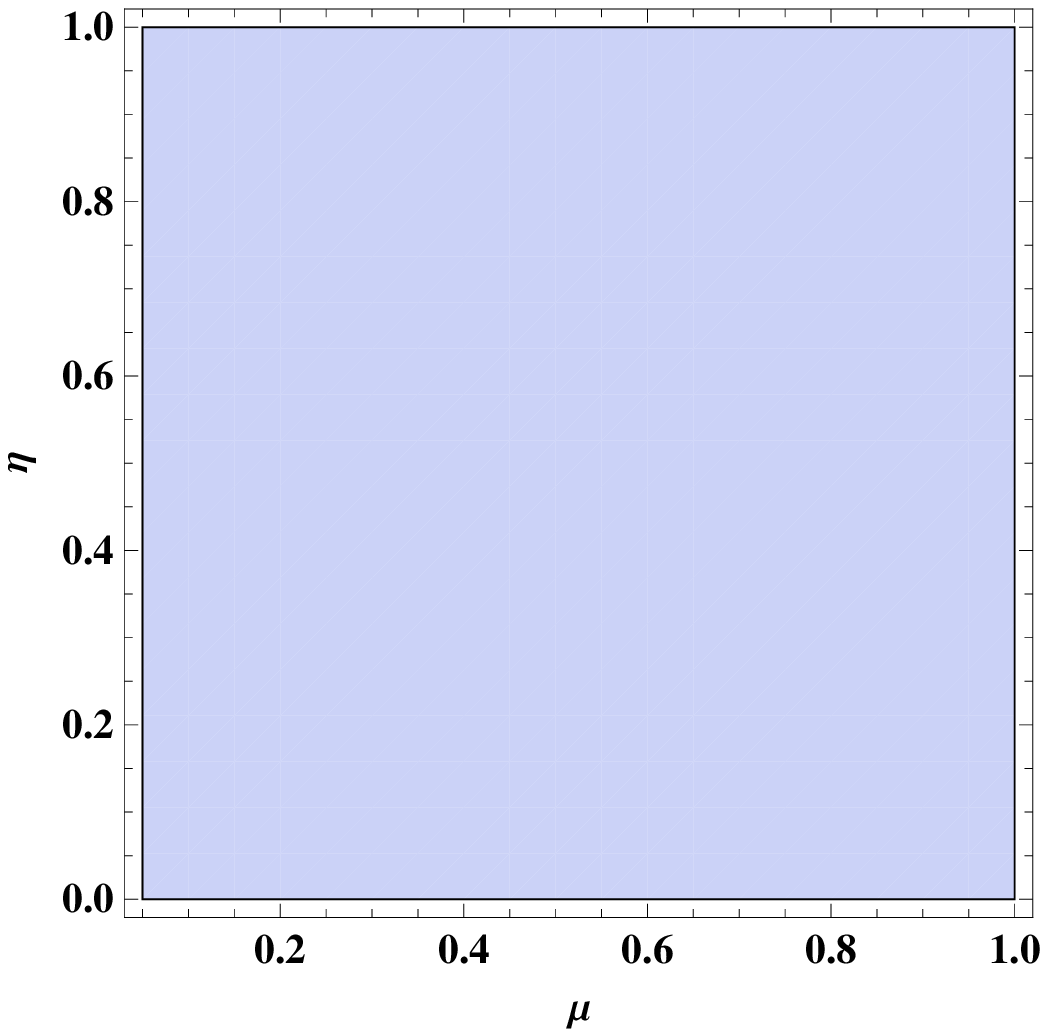}
\caption{(Color online) Security possible for $(\eta,\mu)$ with decoy states in the
  shaded region where \eqref{eq:cond1} is fulfilled. 
Additional conditions such as
  \eqref{eq:cond2} are checked in the following
  Figure~\ref{fig:decoyCond2numu}.}
\label{fig:decoyCond1numu}
\end{center}
\end{minipage}\hspace{5mm}
\begin{minipage}[t]{0.45\textwidth}
\begin{center}
\includegraphics[width=5cm]{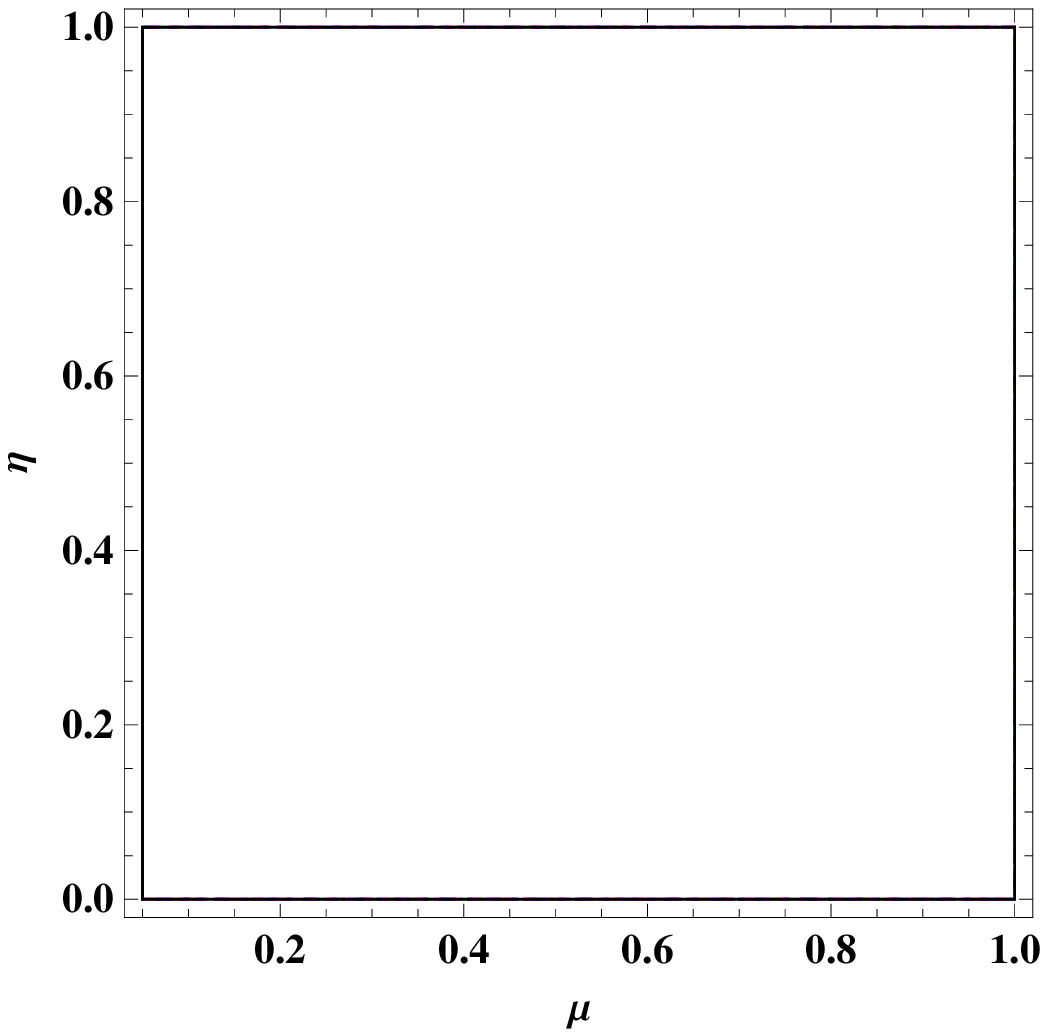}
\caption{(Color online) Security possible for $(\eta,\mu)$ with decoy states in the upper enclosed regions for a low storage noise of $r = 0.9$ and
storage rates $1/2$ (dashed red line), $0.45$ (dotted green line), $0.35$ (dot dashed blue line), $0.25$ (large dashed magenta line),
$0.15$ (solid black line). (satisfying~\eqref{eq:cond1} and~\eqref{eq:cond2}).}
\label{fig:decoyCond2numu}
\end{center}
\end{minipage}
\end{figure}

\begin{figure}[h]
\begin{center}
\includegraphics[width=8cm]{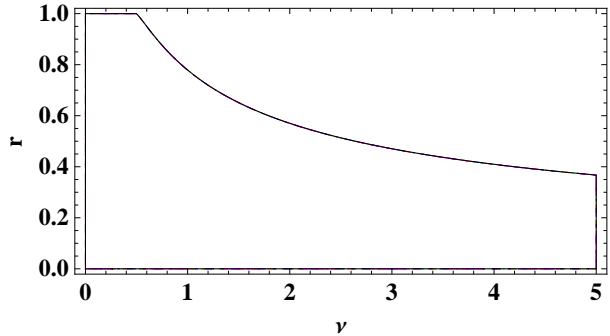}
\caption{(Color online) Security for $(r,\nu)$ with decoy states below
  the lines for $\mu = 0.3$ and detection efficiencies $\eta$: $0.7$
  (solid black line), $0.5$ (large dashed magenta line), $0.4$ (dot
  dashed blue line), $0.3$ (dotted green line), $0.2$ (dashed red
  line).}
\label{fig:decoyNuvsR}
\end{center}
\end{figure}

For direct comparison, we now provide the same plots as given in Section~\ref{sec:wcpWSE1}, where for simplicity we will always choose
$\hat{\mu} = 0.05$. 
Of course, this may not be optimal, but serves as a good comparison.
As expected using decoy states limits dishonest Bob from reporting too many single-photon rounds as missing, thereby allowing us
to place a better bound on $r^{(1)}$.
This fact greatly increases the range of parameters $\eta$ and $\mu$ for which we can hope to show security
as shown in Figures~\ref{fig:decoyCond1numu} and~\ref{fig:decoyCond2numu}.
We also observe in Figure~\ref{fig:decoyNuvsR} that the detection efficiency $\eta$ plays almost no role
in determining for which values of storage noise $r$ and storage rate $\nu$ we can obtain security.
This is true for all values of $\mu \leq 0.4$ we have chosen to examine.

\begin{figure}[h]
\begin{center}
\includegraphics[width=7cm]{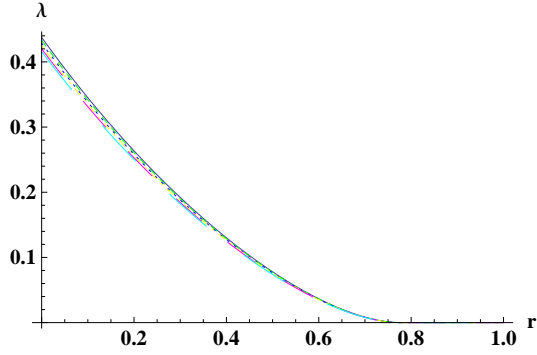}
\caption{(Color online) The WSE rate $\lambda$ for decoy states in terms of the amount of depolarizing noise $r$ where $\mu = 0.3$, and
a variety of detection efficiencies $\eta$: $0.7$ (solid black line), $0.6$ (dashed red line), $0.5$ (dotted blue line),
$0.4$ (dot dashed yellow line), $0.3$ (large dashed magenta line) and $0.2$ (larger dashed turquoise line).}
\label{fig:decoyLambdaExample}
\end{center}
\end{figure}

\begin{figure}[h]
\begin{minipage}[t]{0.45\textwidth}
\begin{center}
\includegraphics[width=5cm]{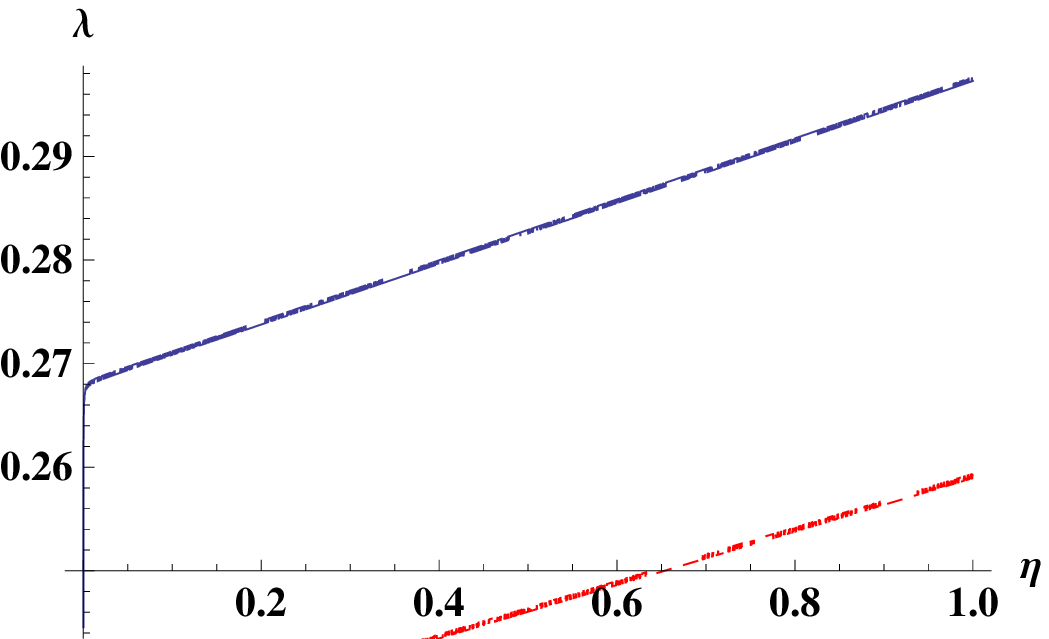}
\caption{(Color online) The WSE rate $\lambda$ for decoy states in terms of the detection efficiency $\eta$ for $r=0.8$ and storage rates
$\nu$: $1/5$ (solid blue line), $1/4$ (dashed red line).} 
\label{fig:decoyLambdavsEta}
\end{center}
\end{minipage}\hspace{5mm}
\begin{minipage}[t]{0.45\textwidth}
\begin{center}
\includegraphics[width=5cm]{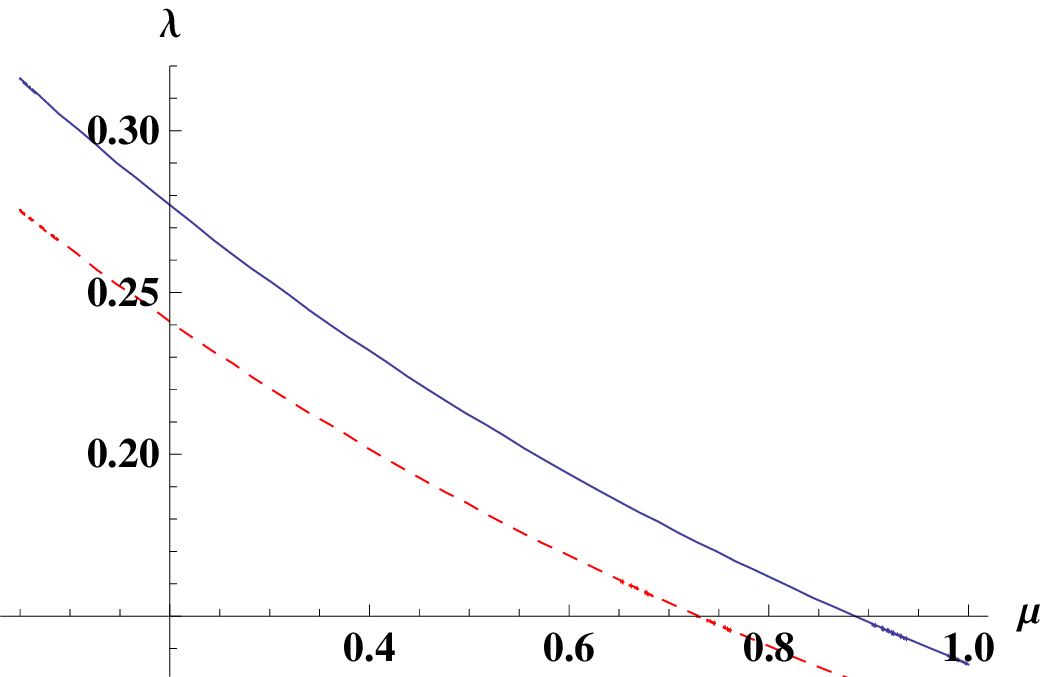}
\caption{(Color online) The WSE rate $\lambda$ for decoy states in terms of the mean photon number $\mu$ for $r=0.8$ and $\nu$ as in Figure~\ref{fig:decoyLambdavsEta}}
\label{fig:decoyLambdavsMu}.
\end{center}
\end{minipage}
\end{figure}

It is however interesting to observe that the magnitude of the final weak string erasure rate $\lambda$ changes only slightly when we use decoy states.
This is due to the strong converse parameter~\eqref{eq:converseParameter} which determines $\lambda$ as given in~\eqref{eq:lambdaOptimize}
and which is not necessarily large for larger values of $R$.
This is witnessed by Figure~\ref{fig:decoyLambdaExample}.
Still, we again observe that we may use much lower values of $\eta$ as shown in Figure~\ref{fig:decoyLambdavsEta} and
a much higher mean photon number $\mu$ as shown in Figure~\ref{fig:decoyLambdavsMu}.

\subsubsection{1-2 oblivious transfer}

\begin{figure}
\begin{center}
\includegraphics[width=5cm]{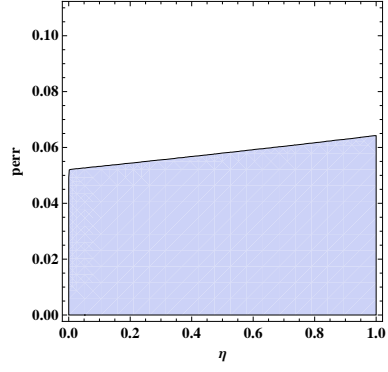}
\caption{(Color online) Security for $(p_{\rm err},\eta)$ with decoy states in the shaded region for example parameters
$r=0.4$, $\nu=1/5$ and large $\omega = 100000$.}
\label{fig:decoyErrvsEta}
\end{center}
\end{figure}

\begin{figure}
\begin{center}
\includegraphics[scale=0.8]{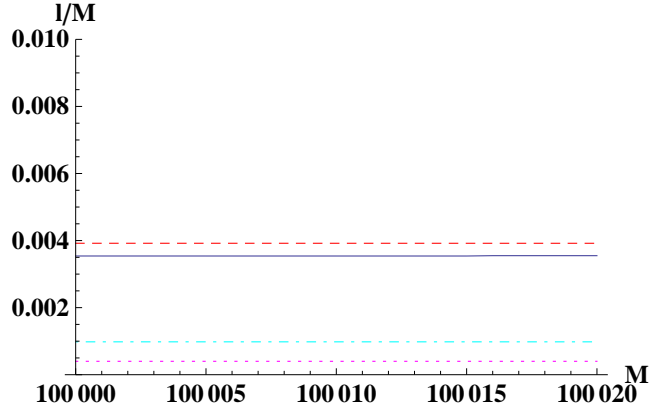}
\caption{(Color online) The rate $\ell/M$ of oblivious transfer with decoy states for a large number of valid pulses $M$ for parameters $\omega=1000$ and
($\mu = 0.2$, $\eta = 0.3$, $r=0.1$, $\nu=1/10$, solid blue line),
($\mu = 0.3$,$\eta=0.7$, $r=0.1$, $\nu=1/10$, dashed red line),
($\mu=0.3$, $\eta=0.7$, $r=0.7$, $\nu=1/4$, dotted magenta line),
($\mu=0.3$, $\eta=0.7$, $r=0.4$, $\nu=1/3$, light blue line).}
\label{fig:decoyRatePlot}
\end{center}
\end{figure}

Again, we also consider the security of $(\ell, \eps)$-oblivious transfer based on weak string erasure implemented using a weak coherent source and
decoy states as above.
We first observe that decoy states soften the trade-off between 
the bit error $p_{\rm err}$ and the efficiency $\eta$ as shown in Figure~\ref{fig:decoyErrvsEta},
where we for now treat $p_{\rm err}$ as an independent parameter to get an intuition for its contribution.
Figure~\ref{fig:decoyRatePlot} now shows how many bits $\ell$ of 1-2 oblivious transfer we can hope to obtain per valid pulse $M$
for very large $M$, when using decoy states. Again, we see that using decoy states softens the effects of $\eta$.
Note that we again count only the valid pulses, which here corresponds to all pulses sent with the signal setting.
As in QKD, it may be possible to use the remaining pulses which one could incorporate in our analysis given in the appendix.
However, for clarity of exposition, we have chosen not to make use of such pulses in this work.

\subsection{Parametric down-conversion source}\label{sec:pdc}

\subsubsection{Experimental setup and loss model}

\begin{figure}
\begin{center}
\includegraphics[scale=0.8]{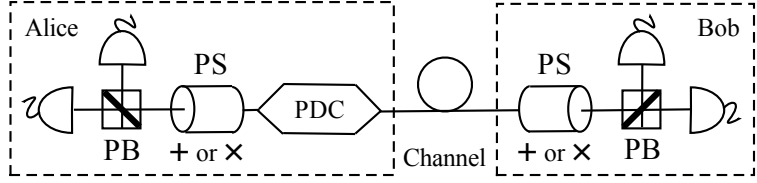}
\end{center}
\caption{Experimental setup with a PDC source. Alice and Bob measure each output signal 
by means of an active BB84 measurement setup, like the one described in Section~\ref{sec_wcp_pract}.}
\label{figure:pdc}
\end{figure}

Now we consider that Alice uses a pumped type-II PDC source. 
The states emitted by this type of source can be 
written as \cite{pdc}
\begin{equation}\label{marcos1}
\ket{\Psi_{\rm src}}_{AB} = \sum_{n=0}^{\infty} \sqrt{p^n_\src} \ket{\Phi_n}_{AB},
\end{equation}
where the probability distribution $p^n_\src$ is given by
\begin{align}
p^n_\src = \frac{(n+1)(\mu/2)^n}{(1+(\mu/2))^{n+2}}.
\end{align}
The parameter $\mu/2$ is directly related to the pump amplitude of the laser resulting in a mean photon pair number per pulse of $\mu$, and
\begin{equation}\label{marcos2}
\ket{\Phi_n}_{AB} = \sum_{m=0}^n \frac{(-1)^m}{\sqrt{n+1}} \ket{n-m,m}_A \ket{m,n-m}_B.
\end{equation}
Here we have used the computational basis on each side. Each signal state $\ket{\Phi_n}_{AB}$
contains exactly $2n$ photons; $n$ of them are measured by Alice and the other $n$ are measured by Bob, as depicted in  Figure~\ref{figure:pdc}. 
We furthermore choose $\eta$ as in the case of a 
weak coherent source. That is, since $\eta_A\eta_{\rm channel}\eta_B=1$ we simply write $\eta = \eta_D$ for both parties.
The dark count rate is again denoted by $\pdark$.

An important difference between the setup using a PDC source and the
one using a weak coherent pulse source, is that Alice herself can
(with some probability) discard a round if she concludes no
photon---or too many photons---have been emitted.  These rounds can be
safely discarded by herself, and thus do not contribute to the
protocol any further.  We will refer to the remaining pulses as
\emph{valid}.  To compare the two approaches more easily, we will
assume that in the case of a PDC source, we consider
only the valid pulses. That is, the parameter $M$ in the WSE protocol
corresponds to the valid pulses, and not to all pulses emitted by
Alice. It is certainly debatable whether this is a fair comparison,
but since $M$ is the parameter which is relevant to the security of
the protocol, we choose to consider the final rates as a function of
$M$. 

The setting of a PDC source is slightly more difficult to analyze, but
can lead to better rates $\ell/M$ than those arising from a weak
coherent source, where, like before, $\ell$ is the number of bits of
oblivious transfer we obtain and $M$ is the number of valid pulses.
The reason for this improvement is two-fold: First, from her
measurement results, Alice can (with some probability) estimate how
many photons have been emitted each given time. This means that we are
no longer restricted to tuning the source such that the number of
multi-photon emissions is too low, but can permit for a larger
variation by relying on Alice to filter out the unwanted events.
Second, a multi-photon emission does not provide dishonest Bob with
full information about the signal state sent by Alice. In this
scenario we need to consider the probability of success for dishonest
Bob when a certain number of photons have been emitted which is given
by Claim~\ref{lem:comparison} in the appendix. Table~\ref{tab:parametersPDC}
again summarizes the probabilities we need to know in order to
evaluate security.  Since some expressions can be rather unwieldy for
the case of a PDC source, we will sometimes refer to the corresponding
equation in the appendix.

\begin{table}
\begin{center}
\begin{tabular}{|l|l|}
\hline
Parameter & Value\\
\hline
$p^1_\src$ & $\mu/(1+(\mu/2))^3$\\
\hline
$p^1_\sent$ & ~\eqref{eq:pdcSent}\\
\hline
$p^{h|1}_{\clickB}$ & $\eta + (1-\eta) \pdark (2 - \pdark)$\\
\hline
$p^{d,n}_{B,\rm err}$ & ~\eqref{mm3}\\
\hline
$p^{d}_\lostB$ & $p^0_\sent$, see~\eqref{eq:pdcSent}\\
\hline
$p^{h}_{\slostB}$ & ~\eqref{mm}\\
\hline
$p^{h}_{\lostB}$ & $p^h_{\slostB} - p^h_{\slostB} \pdark(2-\pdark)$\\
\hline
$p_\derr$ & $\pdark(1-\pdark) + \pdark^2/2$\\
\hline
$p_\dserr$ & ~(\ref{mm2})\\
\hline
$p^h_\serr$ & ~\eqref{eq:pdcSignalErr}\\
\hline
$p^h_{\errB}$ & $p^h_\serr (1 - \pdark (2-\pdark))+ p^h_{\slostB} 
p_\derr + p_\dserr$\\
\hline
\end{tabular}
\end{center}
\caption{Summary of probabilities for parametric down-conversion
  source \label{tab:parametersPDC}}
\end{table}

\subsubsection{Security parameters}

\begin{figure}[h]
\begin{minipage}[t]{0.45\textwidth}
\begin{center}
\includegraphics[width=5cm]{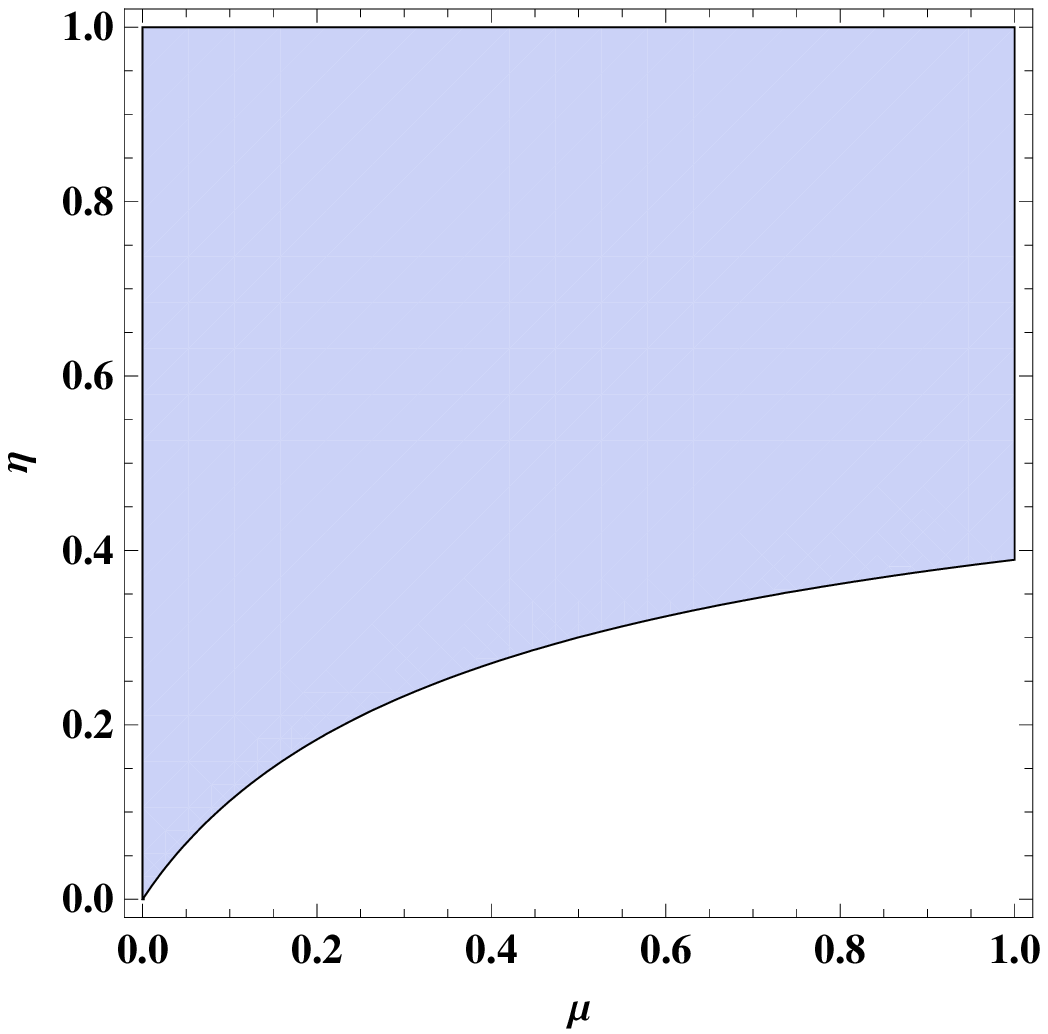}
\caption{(Color online) Security possible for $(\eta,\mu)$ in the shaded region where
\eqref{eq:cond1} is fulfilled. Our proof does not apply to parameters
in the region below the curve. For the shaded region above the curve,
additional conditions such as \eqref{eq:cond2} are checked in the
following Figure~\ref{mm4}.}
\label{fig:pdcCond1numu}
\end{center}
\end{minipage}\hspace{5mm}
\begin{minipage}[t]{0.45\textwidth}
\begin{center}
\includegraphics[width=5cm]{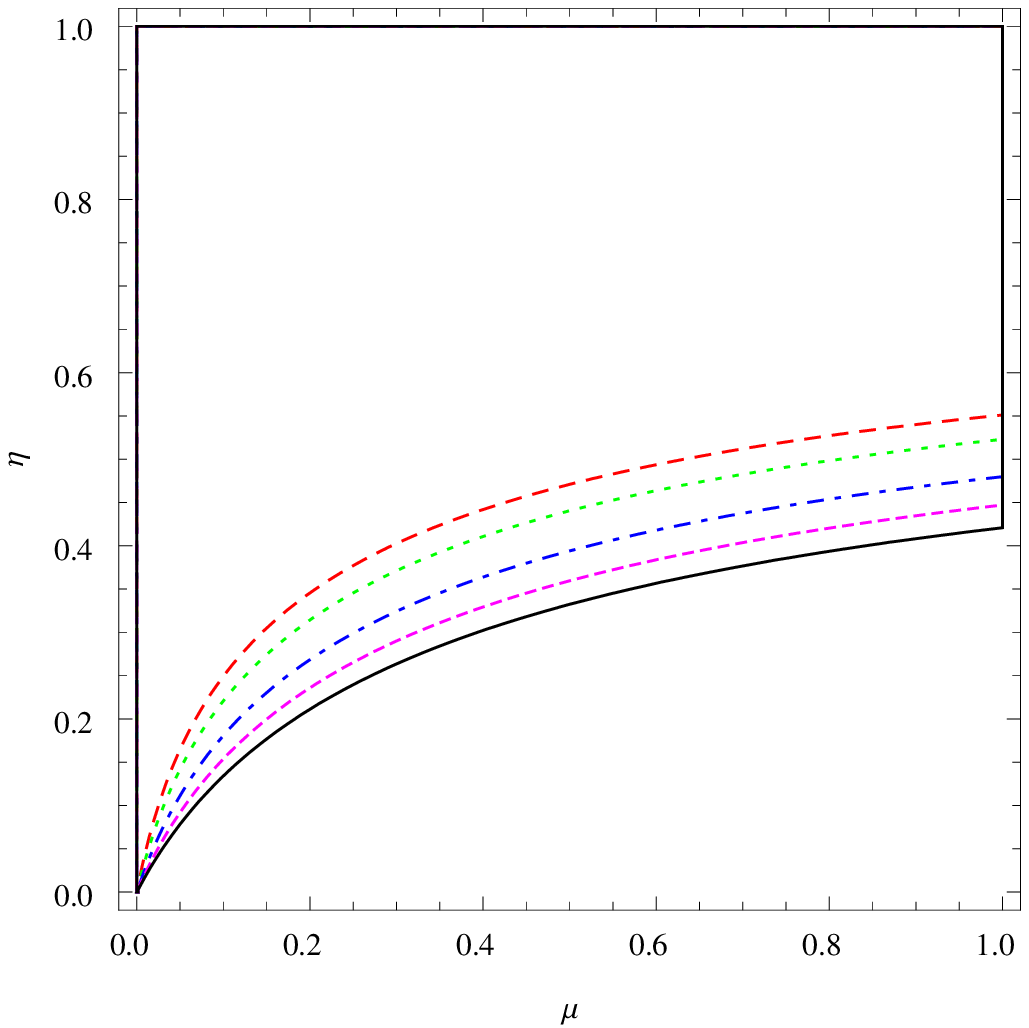}
\caption{(Color online) Security possible for $(\eta,\mu)$ in the upper enclosed regions for a low storage noise of $r = 0.9$ and
storage rates $1/2$ (dashed red line), $0.45$ (dotted green line), $0.35$ (dot dashed blue line), $0.25$ (large dashed magenta line),
$0.15$ (solid black line). (satisfying~\eqref{eq:cond1} and~\eqref{eq:cond2}). \label{mm4}}
\end{center}
\end{minipage}
\end{figure}

\paragraph{Weak string erasure}
We now investigate the security of $(m,\lambda,\eps,p_{\rm err})$-weak string erasure, when using a PDC source.
For easy comparison, we will consider exactly the same plots as before, where however we sometimes choose a different value for the mean photon
number which seemed more useful for this source.
For simplicity, we will also consider a setting where we give all the information encoded in multi-photons to
dishonest Bob for free, {\it i.e.}, we consider $p^{d,n}_{B,\rm err}=0$, which clearly overestimates his capabilities
as we see in Claim~\ref{lem:comparison}.
Again, we first consider when security can
be obtained in principle (i.e., when~\eqref{eq:cond1} and~\eqref{eq:cond2} are satisfied)
as a function of the mean photon number $\mu$, the detection efficiency $\eta$, the storage rate $\nu$
and the amount of storage noise, where our examples here focus on the depolarizing channel with parameter $r$ as defined in~\eqref{eq:depol}.
Figure~\ref{fig:pdcCond1numu} thereby tells 
us again when security is possible at all, independently of the amount of storage noise.
As before, we then examine a particular example of storage noise and storage rates in 
Figure~\ref{mm4}, that even for low storage noise, we can hope
to achieve security for many source settings.

\begin{figure}[h]
\begin{center}
\includegraphics[width=8cm]{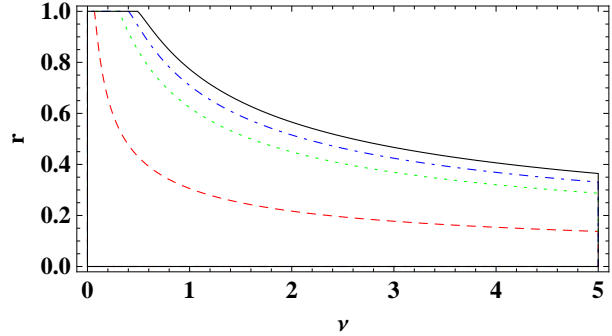}
\caption{(Color online) Security for $(r,\nu)$ below the lines for $\mu = 0.3$ and
detection efficiencies $\eta$: $0.7$ (solid black line), $0.5$ (large dashed magenta line), $0.4$ (dot dashed blue line),
$0.3$ (dotted green line), $0.2$ (dashed red line).}
\label{fig:pdcNuvsR}
\end{center}
\end{figure}

Second, we consider again when conditions~\eqref{eq:cond1} and~\eqref{eq:cond2} can be satisfied in terms of the
amount of noise in storage given by $r$, and the storage rate $\nu$ for some typical parameters in an experimental setup in
Figure~\ref{fig:pdcNuvsR}. It is interesting to note that the efficiency $\eta$ plays a much more prominent role when using a PDC source. This comes from the fact that Alice herself also uses a detector of efficiency $\eta$ to post-select some of
the pulses.

\begin{figure}[h]
\begin{center}
\includegraphics[width=7cm]{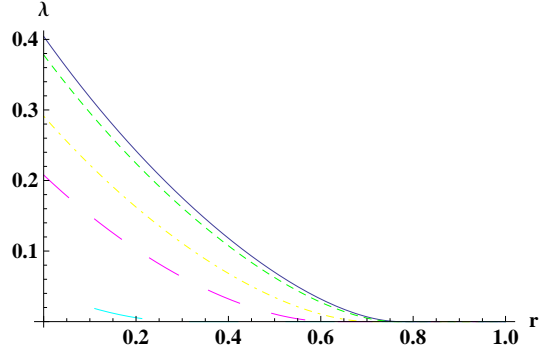}
\caption{(Color online) The WSE rate $\lambda$ in terms of the amount of depolarizing noise $r$ where $\mu = 0.3$, and
a variety of detection efficiencies $\eta$: $0.7$ (solid black line), $0.6$ (dashed red line), $0.5$ (dotted blue line),
$0.4$ (dot dashed yellow line), $0.3$ (large dashed magenta line) and $0.2$ (larger dashed turquoise line).}
\label{fig:pdcLambdaExample}
\end{center}
\end{figure}

Yet, we conclude that secure weak string erasure can be obtained for a reasonable choice of parameters, so it remains to establish
the weak string erasure rate $\lambda$ by solving the optimization problem given by~\eqref{eq:lambdaOptimize}.
Figure~\ref{fig:pdcLambdaExample} gives us $\lambda$
for various choices of experimental settings, and a storage rate of $\nu=1$.
This demonstrates that even for a very high storage rate, there is a positive rate of $\lambda$ for many reasonable settings.

\begin{figure}[h]
\begin{minipage}[t]{0.45\textwidth}
\begin{center}
\includegraphics[width=5cm]{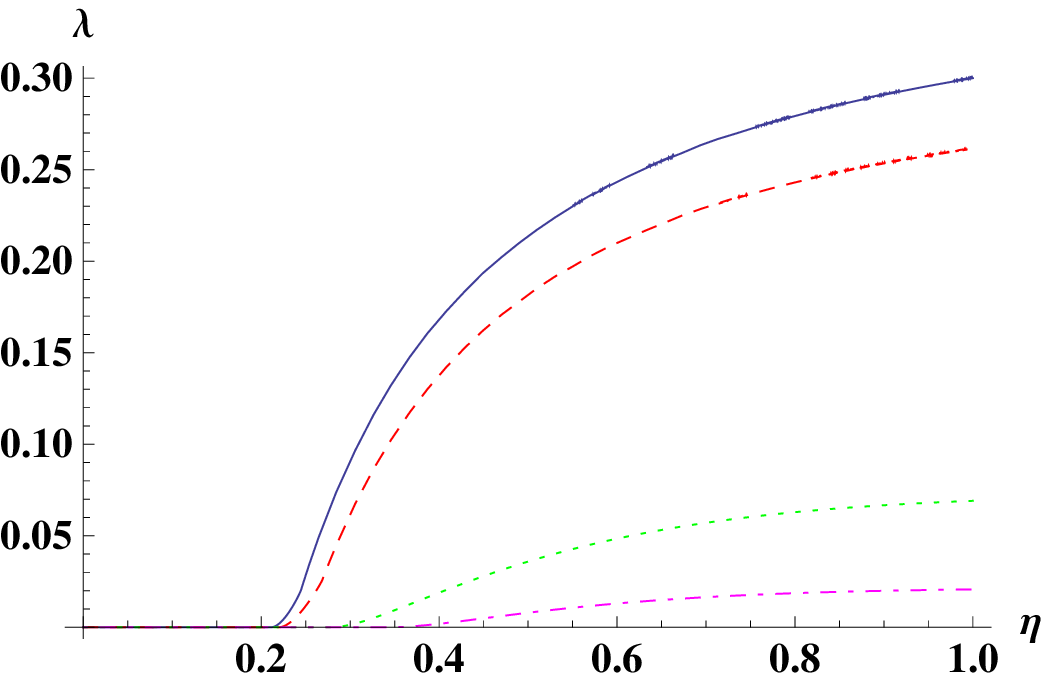}
\caption{(Color online) The WSE rate $\lambda$ in terms of the detection efficiency $\eta$ for $r=0.8$ and storage rates
$\nu$: $1/5$ (solid blue line), $1/4$ (dashed red line), $1/2$ (dotted green line), $2/3$ (dot dashed magenta line).}
\label{fig:pdcLambdavsEta}
\end{center}
\end{minipage}\hspace{5mm}
\begin{minipage}[t]{0.45\textwidth}
\begin{center}
\includegraphics[width=5cm]{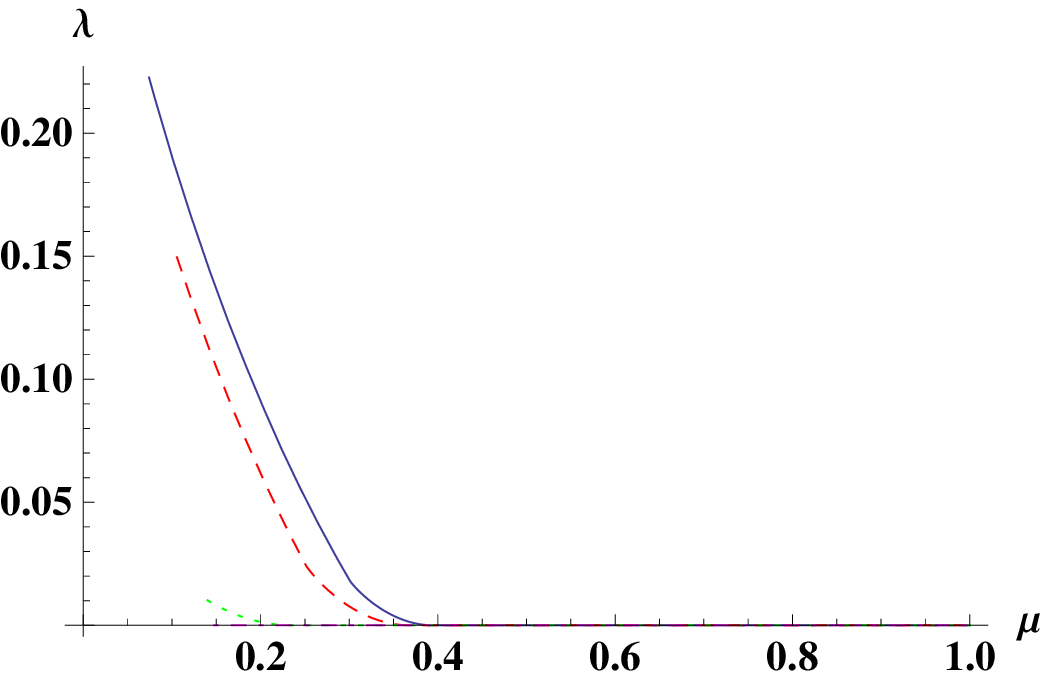}
\caption{(Color online) The WSE rate $\lambda$ in terms of the mean photon number $\mu$ for $r=0.8$ and $\nu$ as in Figure~\ref{fig:pdcLambdavsEta}.}
\label{fig:pdcLambdavsMu}
\end{center}
\end{minipage}
\end{figure}

The trade-off between $\lambda$ and the detection efficiency $\eta$ given in Figure~\ref{fig:pdcLambdavsEta}
is quite similar to what we observed in the case of a weak coherent source.
On the other hand, the trade-off between $\lambda$ and the mean photon number $\mu$ in Figure~\ref{fig:pdcLambdavsMu} shows
that having a low mean photon number seems more significant. Recall, however, that we
have for simplicity assumed that we give all multi-photons to Bob for free which greatly overestimates his capabilities when using 
a PDC source. These parameters could thus be improved when including multi-photons.

\paragraph{1-2 oblivious transfer}\label{sec:pdcOT1}

\begin{figure}
\begin{center}
\includegraphics[width=5cm]{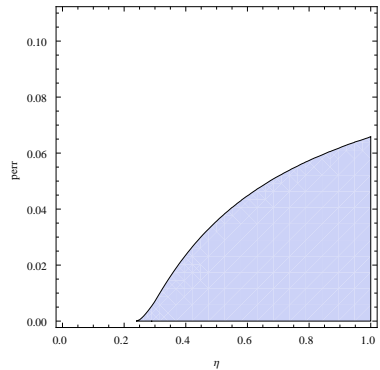}
\caption{(Color online) Security for $(p_{\rm err},\eta)$ in the shaded region for example parameters
$r=0.4$, $\nu=1/5$ and large $\omega = 100000$.}
\label{fig:pdcErrvsEta}
\end{center}
\end{figure}

We can now consider the security of $(\ell, \eps)$-oblivious transfer based on weak string erasure implemented using a PDC source.
In Figure~\ref{fig:pdcErrvsEta}, we first examine the trade-off between an independently chosen bit error rate $p_{\rm err}$ and the efficiency $\eta$,
which is similar to what we observe for the case of a weak coherent source.

\begin{figure}
\begin{center}
\includegraphics[scale=0.8]{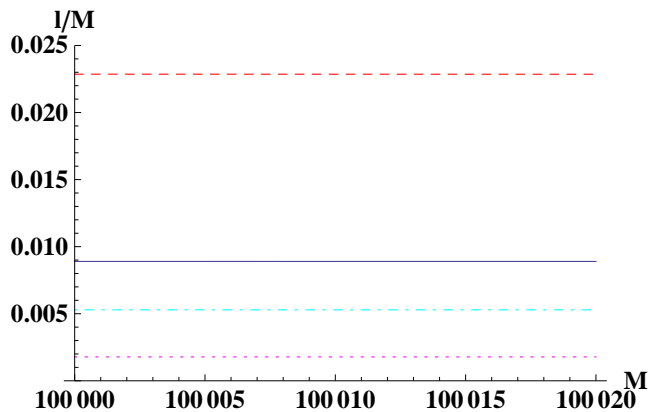}
\caption{(Color online) The rate $\ell/M$ of oblivious transfer for a large number of valid pulses $M$ for parameters $\mu = 0.05$, $\omega=1000$ and
($\eta = 0.3$, $r=0.1$, $\nu=1/10$, solid blue line),
($\eta=0.7$, $r=0.1$, $\nu=1/10$, dashed red line),
($\eta=0.7$, $r=0.7$, $\nu=1/4$, dotted magenta line),
($\eta=0.7$, $r=0.4$, $\nu=1/3$, light blue line)}.
Note that the scaling of this plot is different than for the WCP source with and without decoy states.
\label{fig:pdcRatePlot}
\end{center}
\end{figure}

Figure~\ref{fig:pdcRatePlot} now shows 
how many bits $\ell$ of 1-2 oblivious transfer we can hope to obtain per valid pulse $M$
for very large $M$. 
This is much higher than what we observe for the case of a weak coherent source, but note that in all 
plots we only consider the \emph{valid} pulses $M$. 
For a weak coherent source, this is equal to the actual number of pulses
emitted as Alice does not post-select. However, for the case of a PDC source, Alice can (with some probability) discard rounds
in which no photon has been emitted.
This comparison is arguably unfair, but since $M$ is the parameter that is relevant to the security of our protocol, we chose
to use the number of valid pulses, instead of the number of all pulses.

\section{Conclusions and open questions}

We have shown that security in the noisy-storage
model~\cite{prl:noisy,noisy:new} can in principle be obtained in a
practical setting, and provided explicit security parameters for two
possible experimental setups. Our analysis shows that the protocols
of~\cite{noisy:new} are well within reach of today's technology.

We have been mostly focusing our attention on short-distance (in the
order of a few meters) applications. For this range, it is an
interesting experimental challenge to construct small handheld devices
which can be used to implement these protocols.  Nonetheless, in the
future it might be interesting to study the curve between the rate and
the distance of secure WSEE (in a similar way as the key rate versus
distance curve in QKD). Such a curve will allow us to see if our
protocols can be applied in a local area network (LAN) or metropolitan
area network (MAN). Note that for medium-distance (say order 10km)
applications, our protocol may still work. For instance, standard
telecom fiber has a channel loss of about 0.2dB/km at telecom
wavelength (i.e. 1550nm) So, 10km translates to only 2dB channel loss,
which seems quite manageable!

Many important theoretical (see~\cite{noisy:new}) as well as practical
issues remain to be addressed.  As in quantum key distribution (QKD),
we have assumed that all experimental components behave as we expect
them to. Hence, we have not considered any practical attacks such as
exploiting detectors that are blind above a certain
threshold~\cite{vadim:blinding}, which is outside the scope of this
work.  Most importantly however, it is certaintly possible to improve
the parameters obtained here. These improvements can come from
theoretical advances~\cite{noisy:new}, as well as an exact
optimization of all parameters for a particular experimental setup.
Furthermore, in the case of parametric down conversion, for example,
we have not made use of the fact that Bob cannot gain full information
from multi-photon emissions, which leads to an increase in
rates. Similarly, when using decoy states, one could make use of
pulses emitted using a decoy setting in the protocol. This requires a
careful analysis of weak string erasure for different photon sources
analogous to the one presented in the appendix.  Nevertheless, we hope
that this analysis paves the way for a practical implementation of
protocols in the noisy-storage model.

\acknowledgments

We thank Matthias Christandl, Andrew Doherty, Chris Ervens, Robert
K{\"o}nig, Prabha Mandayam, John Preskill, Joe Renes, Gregor Weihs and
J{\"u}rg Wullschleger for interesting discussions on various aspects
of the noisy-storage model. SW is supported by NSF grants PHY-04056720
and PHY-0803371. MC is supported by
Xunta de Galicia (Spain, Grant No. INCITE08PXIB322257PR). 
CS is supported by the EU fifth framework project QAP
IST 015848 and the NWO VICI project 2004-2009. 
HL is supported by funding agencies CFI, CIPI, the CRC program, CIFAR, MITACS, NSERC, OIT and QuantumWorks.
SW and HL also thank the KITP program on quantum information science funded by NSF grant PHY-0551164.
Part of this work was
carried out during visits to the University of Toronto, Caltech and KITP. We thank these institutions for their kind hospitality.

\bibliographystyle{apsrev}

\appendix

\section{Proof of security: WSEE} \label{app:wsee} Here we show how
the security proof of~\cite{noisy:new} can be modified to apply in the
practical settings considered in this paper. To this end, we first provide a
more formal definition of WSEE.

\begin{definition}\label{def:wseWithErrors}
An {\em $(m,\lambda,\varepsilon,p_{\rm err})$-weak string erasure protocol with errors} (WSEE) is a protocol between Alice and Bob satisfying the following properties, where $\cE_{p_{\rm err}}$ is defined as in~\eqref{eq:bitErrors}:
\begin{description}
\item[Correctness:]
If both parties are honest, then the ideal state $\sigma_{X^m \cI \cE_{p_{\rm err}}(X_{\cI})}$ is defined such that
\begin{enumerate}
\item The joint distribution of the $m$-bit string $X^m$ and subset $\cI$ is uniform:
\begin{align}
\sigma_{X^m\cI}&=\tau_{\sbin^m}\otimes\tau_{2^{[m]}}\ ,
\end{align}
\item The joint state $\rho_{AB}$ created by the real protocol is $\eps$-close to the ideal state:
\begin{align}
\rho_{AB} \approx_\eps \sigma_{X^m\cI \cE_{p_{\rm err}}(X_\cI)}\ .\label{eq:strongerasure}
\end{align}
where we identify $(A,B)$ with $(X^m,\cI \cE_{p_{\rm err}}(X_{\cI}))$.
\end{enumerate}
\item[Security for Alice:] If Alice is honest, then there exists an ideal state
$\sigma_{X^mB'}$ such that
\begin{enumerate}
\item The amount of information $B'$ gives Bob about $X^m$ is limited:
\begin{align}
\frac{1}{m}\hmin(X^m|B')_\sigma\geq \lambda
\end{align}
\item
The joint state $\rho_{AB'}$ created by the real protocol is $\eps$-close to the ideal state:
\begin{align}
\sigma_{X^mB'} \approx_\eps \rho_{AB'}
\end{align}
where we identify $(X^m,B')$ with $(A,B')$.
\end{enumerate}

\item[Security for Bob:] If Bob is honest,
then there exists an ideal state $\sigma_{A'\widehat{X}^m\cI}\,$, where $\widehat{X}^m \in \01^m$ and $\cI \subseteq [m]$ such that
\begin{enumerate}
\item The random variable $\cI$ is independent of $A'\widehat{X}^m$ and uniformly distributed over $2^{[m]}$:
\begin{align}
\sigma_{A'\widehat{X}^m\cI}=\sigma_{A'\widehat{X}^m}\otimes\tau_{2^{[m]}}\ .
\end{align}
\item
The joint state $\rho_{A'B}$ created by the real protocol is $\eps$-close to the ideal state:
\begin{align}
\rho_{A'B} \approx_\eps \sigma_{A'\cI \cE_{p_{\rm err}}(\widehat{X}_\cI)}
\end{align}
where we identify $(A',B)$ with $(A',\cI \cE_{p_{\rm err}}(\widehat{X}_\cI))$.
\end{enumerate}
\end{description}
\end{definition}

We study Protocol 1, i.e., without the use of decoy states.  The case
of decoy states is analogous, where we obtain a different bound
in~\eqref{eq:r1Bound}, as discussed in Section~\ref{sec:proofDecoy}.
The analysis is essentially the same in both cases, only we bound
certain parameters in a different way.  The general security
evaluation of correctness and the case when Bob is honest follows the
same arguments as in~\cite{noisy:new}. It is clear by construction that
an honest Bob reports enough rounds so that Alice does not abort
except with probability $\eps$ and hence the real output states are at most
$\eps$-far from the ideal states.

From now on we concentrate on the situation where Alice is honest, but
Bob might try to cheat.  Our analysis contains two steps. We first
consider single-photon emissions, which we analyze as
in~\cite{noisy:new}, taking into account that Bob may report some
additional single-photon rounds as missing. Second, we consider
multi-photon rounds. The main difficulty arises from the fact that Bob
may report up to
\begin{align}
M^d_{\rm max} = (p^h_\lostB + \zeta^h_\lostB)M
\end{align}
of the $M$ rounds as missing, where he himself can choose which rounds to report. First of all, note that we 
can assume that even a dishonest Bob always reports a round as missing if he receives a vacuum state. By the same arguments as in Section~\ref{sec:parameters} we have that
the number of rounds where Bob observes no click lies in the interval 
$[(p^d_\lostB - \zeta^d_\lostB)M, (p^d_\lostB + \zeta^d_\lostB)M]$
for $\zeta_\lostB^d = \sqrt{\ln(2/\eps)/(2M)}$, except with probability $\eps$. 
Here, we make a worst-case assumption that the number of rounds where dishonest Bob observes no click is
given by
\begin{align}
M^d_{\rm nc} = (p^d_\lostB - \zeta^d_\lostB)M\ ,
\end{align}
and he can thus report up to
\begin{align}
M^d_{\rm report} = M^d_{\rm max} - M^d_{\rm nc} = (p^h_\lostB - p^d_\lostB + \zeta^h_\lostB + \zeta^d_\lostB)M\ ,
\end{align}
rounds of his choice to be missing. Let $M^{(n)}$ denote the number of rounds
corresponding to an $n$-photon emission, let $r^{(n)}$ denote the fraction of $n$ photon 
rounds that dishonest Bob chooses to report as missing, and let
$M^{(n)}_{\rm left} = (1-r^{(n)}) M^{(n)}$ denote the number of $n$ photon
rounds that dishonest Bob has left. Note that in the limit of large $M$ we have
$M^{(n)} = p^n_{\rm sent} M$.
Clearly, we must have that
\begin{align}\label{eq:r1BoundWithM}
\sum_{n=1}^{\infty} r^{(n)} M^{(n)} \leq M^d_{\rm report}\ ,
\end{align}
or Alice will abort the protocol.

\subsection{Single-photon emissions}\label{sec:singlePhoton}

Single photons are desirable, since they correspond to the idealized
setting analyzed in~\cite{noisy:new} where Alice does indeed send BB84
states. Clearly, in the limit of large $M$, we expect roughly
$p^1_{\rm sent} M$ single-photon rounds. However, since Bob may choose
to report single-photon rounds as missing, we have to analyze how many
rounds still contribute to our security analysis. The analysis
of~\cite{noisy:new} links the security to the rate at which Bob has to
send classical information through his noisy storage channel. In order
to determine this rate, we first investigate the setting where he is
not allowed to keep any quantum state.

Let $X^{(1)}$ denote the substring of $X^M$ that corresponds to
single-photon emissions. In~\cite{noisy:new} the rate at which Bob
needs to send information through his noisy-storage channel depends
on an uncertainty relation using post-measurement information. This
uncertainty relation provides a bound on the min-entropy that Bob has
about $X^{(1)}$ given a classical measurement outcome $K$, and the
basis information he obtains later on. We are thus interested in the
min-entropy
\begin{align}
\hmin(X^{(1)}|K^{(1)} \Theta^{(1)})_\rho = - \log P_{\rm guess}(X^{(1)}|K^{(1)} \Theta^{(1)})\ ,
\end{align}
where we use $K^{(1)}$ and $\Theta^{(1)}$ to denote Bob's classical
information and the basis information corresponding to the single-photon rounds respectively and
$P_{\rm guess}$ is the probability that Bob guesses the string $X^{(1)}$ maximized over all choices
of measurements anticipating his post-measurement information $\Theta^{(1)}$~\cite{ww:pistar}.
Important for us is the fact that since Alice picks one of the four BB84 encodings uniformly at random
in each time slot, the initial state 
\begin{align}
\rho_{X^{(1)}Q^{(1)}\Theta^{(1)}}
 = \bigotimes_{j=1}^{M^{(1)}} 
\rho_{X^{(1)}_j Q^{(1)}_j \Theta^{(1)}_j}\ ,
\end{align}
has tensor-product form, and it follows from~\cite{ww:pistar} together with~\cite{prl:noisy} that
also the state
\begin{align}
\rho_{X^{(1)}K^{(1)}\Theta^{(1)}}
 = \bigotimes_{j=1}^{M^{(1)}} 
\rho_{X^{(1)}_j K^{(1)}_j \Theta^{(1)}_j}\ ,
\end{align}
is a tensor product, that is, Bob's best strategy to guess $X^{(1)}$ purely with the help of 
classical information $K^{(1)}$ has tensor-product form. It is important to note that this does not mean that Bob does indeed perform a tensor-product attack in general. It merely states that with respect to the uncertainty he has about $X^{(1)}$ given only his classical information and
the basis information if he kept no quantum computation, his best attack would be a tensor-product attack. And hence for any other classical information
that he may obtain from his actual attack in the protocol, this uncertainty is only going to be greater.

We can now use the fact that the min-entropy of a tensor-product state is additive~\cite{prl:noisy}, to conclude that
the min-entropy that Bob has about $X^{(1)}$ given $K^{(1)}$ and $\Theta^{(1)}$ is thus a min-entropy \emph{per bit},
which allows us to compute the remaining min-entropy if Bob reports some of the single-photon
rounds as missing. More precisely, if $X^{(1)}_{\rm left}$ is the substring of $X^M$ corresponding
to the single-photon rounds that Bob does not report as missing, we know from 
the uncertainty relation of~\cite{serge:new} and a purification argument that
\begin{align}\label{eq:urRate}
\hmin^{\hat{\eps}}(X^{(1)}_{\rm left}|K^{(1)}_{\rm left} \Theta^{(1)}_{\rm left}) \geq \left(\frac{1}{2} - 2\delta\right) M^{(1)}_{\rm left}\ ,
\end{align}
where $K^{(1)}_{\rm left}$ and $\Theta^{(1)}_{\rm left}$ correspond to the classical and basis information respectively
for the remaining single-photon rounds, and
\begin{align}
\hat{\eps} = \exp\left(- \frac{\delta^2 M^{(1)}_{\rm left}}{32(2 + \log\frac{1}{\delta})^2}\right)\ .\label{eq:epsInUncertainty}
\end{align}

To determine the security as a whole, we of course need to take into account that dishonest Bob also holds some
quantum information about $X^{(1)}_{\rm left}$, besides his classical information.
We adopt the notation of~\cite{noisy:new} and write
\begin{align}
\rho_{X^{(1)}_{\rm left}
\Theta^{(1)}_{\rm left} K^{(1)} \cF(Q_{in})}&=\frac{1}{(2^{|M^{(1)}_{\rm left}|})^2}
\sum_{\substack{x,\,\theta \\ k \in \cK}}
P_{K|X=x,\Theta=\theta}(k)
\underbrace{\proj{x}}_{\rm Alice}\otimes\underbrace{\proj{\theta}\otimes\proj{k}\otimes \cF(\zeta_{x\theta k})}_{{\rm Bob\ } B^{(1)}}\ ,\label{eq:statetoanalyze}
\end{align}
where Bob holds $B^{(1)} = \Theta^{(1)}_{\rm left} K^{(1)}
\cF(Q_{in})$, and $\zeta_{x\theta k} \in \bop(Q_{in})$ is the state
entering Bob's quantum storage when Alice chose $x$ and $\theta$, and
Bob already extracted some classical information $k$.  Here, $K^{(1)}$
includes all of Bob's classical information, and depending on Bob's
attack may not have tensor-product form.  Nevertheless, we know
from~\cite{noisy:new} that~\eqref{eq:urRate} tells us at which rate
cheating Bob has to send information through his storage channel $\cF$
for any attack he conceives.

\subsubsection{General storage noise}

In particular, we can now make use of the uncertainty
relation~\eqref{eq:urRate} together with the analysis of~\cite[Lemma
2.2 and Theorem 3.3]{noisy:new} to obtain that for single-photon
rounds we have that for any attack of dishonest Bob
\begin{align}
\hmin^{\eps}(X^{(1)}_{\rm left}|\Theta^{(1)}_{\rm left} K^{(1)}_{\rm left} \cF(Q^{(1)}))
\geq 
- 
\log P_{\rm succ}^{\cF}\left(\left(\frac{1}{2} - \delta\right) M^{(1)}_{\rm left}\right)\ ,
\end{align}
for 
\begin{align}
\eps = 2 \exp\left(
- \frac{(\delta/4)^2}{32(2 + \log(4/\delta))^2} \cdot M^{(1)}_{\rm left}\right).
\end{align}
Note that we have from~\eqref{eq:r1BoundWithM}
that
\begin{align}\label{eq:r1Bound}
r^{(1)} \leq \min\left[\frac{p^h_{\lostB} - p^d_{\lostB} + \zeta^h_\lostB + \zeta^d_\lostB}{p^1_\sent - \zeta^1_\sent},1\right]\ ,
\end{align}
and hence
\begin{align}
M^{(1)}_{\rm left} &= (1 - r^{(1)}) M^{(1)}\\
&\geq 
M \left(1 - \frac{p^h_\lostB - p^d_\lostB +\zeta^h_\lostB + \zeta^d_\lostB}
{p^1_\sent - \zeta^1_\sent}\right)(p^1_{\sent} - \zeta^h_{\sent})\ ,
\end{align}
which in the limit of large $M$ gives us
\begin{align}
M^{(1)}_{\rm left} \geq M (p^1_\sent + p^d_\lostB - p^h_\lostB)\ .
\end{align}
Since $r^{(1)}$ is chosen by dishonest Bob and hence is unknown to
Alice, we bound $\eps$ for any strategy of dishonest Bob as
\begin{align}\label{eq:epsBoundTool}
\eps \leq 2 \exp\left\{ 
- \frac{(\delta/4)^2}{32(2 + \log(4/\delta))^2} \cdot 
\left[1 - \left(\frac{p^h_\lostB - p^d_\lostB + \zeta^h_\lostB + \zeta^d_\lostB}{p^1_\sent - \zeta^1_\sent}\right)\right] (p^1_{\sent} - \zeta^h_{\sent})M  
\right\}\ .
\end{align}
In the case of decoy states, we just obtain a better bound in~\eqref{eq:r1Bound}, where the remaining security analysis is analogous.

\subsubsection{Tensor-product channels} 

Of particular interest is the case where Bob's storage noise is of the
form $\cF = \cN^{\otimes \nu M_{\rm store}}$ where $\nu$ is the
storage rate, $M_{\rm store}$ is the number of bits we count to
determine Bob's storage, and $\cN$ obeys the strong converse
property~\cite{rs:converse}. As outlined earlier, we assume that
the number of qubits that determines Bob's storage size is as in the
idealistic setting of~\cite{noisy:new} given by the number of
single-photon emissions that we expect an honest Bob to receive for
large $M$, {\it i.e.}, $M_{\rm store} $.

From the strong converse property of $\cN$ follows that
\begin{align}
- \log P_{\rm succ}^{\cN^{\otimes \nu M_{\rm store}}}(M_{\rm store} R) \geq \nu \cdot \gamma^{\cN}(R/\nu) M_{\rm store}\ ,
\end{align}
where $\gamma^{\cN}(R/\nu) > 0$ for $C_\cN \cdot \nu < R$ and
$C_\cN$ is the classical capacity of the channel
$\cN$~\cite{rs:converse}.  To achieve security in this setting we
hence want to determine $R$ such that
\begin{align}
\left(\frac{1}{2}-\delta\right) M^{(1)}_{\rm left} = R \cdot M_{\rm store}\ ,
\end{align}
which gives us
\begin{align}\label{eq:defR}
R = \left(\frac{1}{2} - \delta\right) \frac{(1 - r^{(1)})(p^1_\sent -
  \zeta^1_\sent)}{p^1_\sent \cdot p^{h|1}_\clickB} \mbox{ for }
p^{h|1}_\clickB > 0\ ,
\end{align}
and $R = 0$ otherwise, which for large $M$ becomes
\begin{align}
R = \left(\frac{1}{2} - \delta\right) \frac{1 -
  r^{(1)}}{p^{h|1}_\clickB}\ .
\end{align}
Whenever $p^{h|1}_\clickB > 0$,
note that $R$ can be significantly larger than $1/2$ due the difference between $M^{(1)}$ and $M_{\rm store}$.
We can now use~\eqref{eq:r1Bound} to bound $R$
as
\begin{align}\label{eq:boundRatefinite}
R \geq \left(\frac{1}{2} - \delta\right) 
\max\left[0,
\frac{1}{p^{h|1}_\clickB} 
- 
\frac{p^{h}_\lostB - p^d_\lostB + \zeta^h_\lostB +
  \zeta^d_\lostB}{p^1_{\rm sent} \cdot p^{h|1}_\clickB}\right]\ ,
\end{align}
which for large $M$ is just
\begin{align}\label{eq:boundRate}
R \geq \left(\frac{1}{2} - \delta\right) 
\max\left[0,\frac{p^{1}_{\rm sent} - p^{h}_\lostB +
    p^d_\lostB}{p^1_{\rm sent} \cdot p^{h|1}_\clickB}\right]\ .
\end{align}
Summarizing, we have that for any strategy of dishonest Bob
\begin{align}
\hmin^{\eps}(X^{(1)}_{\rm left}|\Theta^{(1)}_{\rm left} K^{(1)}_{\rm left} \cF(Q^{(1)}))
\geq 
\nu \cdot \gamma^{\cN}\left(\frac{R}{\nu}\right) M_{\rm store}\ .
\end{align}

\subsection{Multi-photon emissions}\label{sec:multiPhoton}

It remains to address the case of multi-photon emissions. We analyze
here a conservative scenario where dishonest Bob obtains the basis
information for free whenever a multi-photon emission occurred. This situation
can only make dishonest Bob more powerful. Note that this also means that Bob
will never attempt to store such emissions, since he will never obtain
more information about them as he already has. We thus assume that Bob
keeps no quantum knowledge about the rounds corresponding to
multi-photon emissions.  We will see below that for the case of a PDC
source, Bob nevertheless does not obtain full information about a bit
in the case of a multi-photon emission.

For an $n$-photon emission, the probability that Bob performs a correct decoding is
given by $(1 - p^{d,n}_{\errB})$. If bit $j$ of $X^M$ was generated by an $N=n$-photon emission, 
we thus have
\begin{align}
\hmin(X_j|\Theta_j K_j \ N=n ) = - \log \left(1 - p^{d,n}_{\errB}\right)\ .
\end{align}
Since we assume that Bob keeps no quantum information about the
multi-photon rounds we may write his state corresponding to the rounds
in which $n>1$ photons have been emitted as
\begin{align}
\rho_{X^{(n)}_{\rm left} B^{(n)}} = \bigotimes_j \rho_{X^{(n)}_{\rm left, j} B^{(n)}_j}\ ,
\end{align}
where $B^{(n)}_j$ is a classical register.
Using the fact that the min-entropy is additive for a tensor-product state~\cite{prl:noisy}, 
we have that
Bob's min-entropy about the substring $X^{(n)}$ of $X^M$ (belonging to $N=n$ photon emissions
that Bob does not report as missing) is given by
\begin{align}
\frac{1}{M^{(n)}_{\rm left}} \hmin(X^{(n)}_{\rm left}|\Theta^{(n)}_{\rm left} 
K^{(n)}_{\rm left} N=n) = - \log \left(1 - p^{d,n}_{\errB}\right).
\end{align}

\subsection{Putting things together}

Let $X^m$ be the substring of bits of $X^M$ that Bob does not report
as missing.  In order to determine the overall security parameters, we
need to determine how much min-entropy dishonest Bob has about
\begin{align}
X^m = \bigcup_{n=1}^{\infty} X^{(n)}_{\rm left}\ .
\end{align}
Since we assume that Bob keeps no quantum information about the multi-photon rounds 
we may write the state of the
system if Bob is dishonest as
\begin{align}\label{eq:BobsTotalState}
\rho_{X^m B'} \isomorph \bigotimes_{n=1}^{\infty} \rho_{X^{(n)}_{\rm left} B^{(n)}}\ ,
\end{align}
where $B^{(n)}$ contains a copy of all classical information available
to Bob, and where we have reordered the systems into parts belonging to
different photon number $n$.  The following theorem comes
from~\cite[Theorem 3.3]{noisy:new}, together with the discussion given
above.
\begin{theorem}[Security against Bob]
Fix $\delta \in ]0,\frac{1}{2}[$ and let
\begin{align}\label{eq:defEps}
\eps = 2 \exp\left(- \frac{(\delta/4)^2}{32 (2 + \log(4/\delta))^2} \cdot M^{(1)}_{\rm left}\right)\ .
\end{align}
Then for any attack of a dishonest Bob with storage $\cF: \bop(\hin) \rightarrow \bop(\hout)$, there
exists a cq-state $\sigma_{X^m B'}$ such that
\begin{enumerate}
\item $\sigma_{X^m B'} \approx_{2\eps} \rho_{X^m B'}$\ ,
\item $\frac{1}{m} \hmin(X^m|B)_\sigma \geq - \frac{1}{m} \left[\log P_{\rm succ}^{\cF}(R \cdot M_{\rm store}) 
+ \sum_{n = 2}^{\infty} M^{(n)}_{\rm left} \log \left(1 - p^{d,n}_{\errB}\right)\right]$\ ,
\end{enumerate}
where $\rho_{X^mB'}$ is given by~\eqref{eq:BobsTotalState}.
\end{theorem}
\begin{proof}
  Let $\sigma_{X^{(1)}_{\rm left} B^{(1)}}$ be defined as in the
  analysis of single-photon emissions in~\cite{noisy:new}. Following
  the same arguments as in~\cite{noisy:new} and adding another $\eps$
  for the probability that the number of rounds in which Bob observes
  no click lies outside the interval $[(p^d_\lostB - \zeta^d_\lostB)M,
  (p^d_\lostB + \zeta^d_\lostB)M]$, we get
  $\frac{1}{2}\|\rho_{X^{(1)}_{\rm left} B^{(1)}} -
  \sigma_{X^{(1)}_{\rm left} B^{(1)}}\|_1 \leq 2 \eps$. Furthermore,
  let $\sigma_{X^{(n)}_{\rm left} B^{(n)}} = \rho_{X^{(n)}_{\rm left}
    B^{(n)}}$ for $n>1$ and let
\begin{align}
\sigma_{X^m B'} = \bigotimes_{n=1}^{\infty} \sigma_{X^{(n)}_{\rm left} B^{(n)}}\ .
\end{align}
Note that by the subadditivity of the trace distance, we have
\begin{align}
\frac{1}{2}\| \rho_{X^m B'} - \sigma_{X^m B'} \|_1 \leq 2 \eps\ .
\end{align}
It remains to show that $\sigma_{X^m B'}$ has high min-entropy.
Note that
\begin{align}
\hmin(X^m|B)_\sigma = \hmin(X^{(1)}_{\rm left}|B^{(1)})_\sigma + 
\sum_{n=2}^{\infty} \hmin(X^{(n)}_{\rm left}|B^{(n)})_\sigma\ ,
\end{align}
where we have used the additivity of the min-entropy for tensor-product states~\cite{prl:noisy}, and that 
conditioning on independent information does not change
the min-entropy. Our claim now follows immediately from Sections~\ref{sec:singlePhoton} and~\ref{sec:multiPhoton}.
\end{proof}

We can again specialize this result to the case of tensor-product channels.
\begin{corollary}[Security against Bob]
Let Bob's storage be described by $\cF = \cN^{\otimes \nu M_{\rm store}}$ with $\nu > 0$, $\cN$ satisfying the
strong converse property~\cite{rs:converse}, and
\begin{align}
C_{\cN} \cdot \nu < \min_{r^{(1)}} R\ ,
\end{align}
where $R$ is defined in \eqref{eq:defR}. 
Fix $\delta \in ]0,\min_{r^{(1)}} R - C_{\cN} \cdot \nu [$.
Then, for any attack of dishonest Bob there exists a cq-state $\sigma_{X^m B'}$ such that
\begin{enumerate}
\item $\sigma_{X^m B'} \approx_{2\eps} \rho_{X^m B'}$\ ,
\item $\frac{1}{m} \hmin(X^m|B')_\sigma \geq \frac{1}{m}\left[M_{\rm store}
\nu \cdot \gamma^{\cN}(R/\nu) - \sum_{n = 2}^{\infty} M^{(n)}_{\rm left} \log \left(1 - p^{d,n}_{\errB}\right)\right]$\ ,
\end{enumerate}
with $\rho_{X^n B'}$
and $\eps$ 
given by \eqref{eq:BobsTotalState} and \eqref{eq:defEps} respectively.
\end{corollary}

Our main theorem now follows by allowing Bob to choose $\{r^{(n)}\}$ minimizing his total min-entropy.
To be able to give an exact security guarantee we bound the parameter $\eps$ which may depend
on dishonest Bob's choice of $r^{(1)}$ using~\eqref{eq:epsBoundTool}.

\begin{theorem}[Weak string erasure]\label{thm:wsemainclaim}
Protocol~1 is an $(m,\lambda(\delta),\eps(\delta),p^h_{\errB})$-weak string erasure protocol for the following
two settings:
\begin{enumerate}
\item
Let Bob's storage be given by $\cF: \bop(\hin) \rightarrow \bop(\hout)$, 
and let $\delta\in ]0,\frac{1}{2}[$. Then we obtain a 
min-entropy rate
\begin{align}
\lambda(\delta)&=
 \min_{\{r^{(n)}\}_n}
\lim_{m\rightarrow \infty}
\frac{1}{m} 
\left[-\log P^\cF_{succ}\left(R \cdot M_{\rm store}\right) - \sum_{n=2}^{\infty} M^{(n)}_{\rm left} \log\left(1 - p^{d,n}_{\errB}\right)\right]\ ,\\
\end{align}
where the minimization is taken over all $\{r^{(n)}\}_n$ such that 
$\sum_{n=1}^{\infty} r^{(n)} M^{(n)} \leq M^d_{\rm report}$ and
\begin{align}
m &= \sum_{n=1}^{\infty} M^{(n)}_{\rm left}\ ,\qquad \qquad M_{\rm
  store} = p^1_{\sent} \cdot p^{h|1}_{\clickB} M\\
R & = \left(\frac{1}{2} - \delta\right) \frac{1-r^{(1)}}{p^{h|1}_{\clickB}}\ ,
\end{align}
and error
\begin{align}
\eps(\delta) \leq 4 \exp\left( 
- \frac{\delta^2}{512(4 + \log\frac{1}{\delta})^2} \cdot 
\left(p^1_\sent - \left(\frac{p^1_\sent(p^h_\lostB - p^d_\lostB + \zeta^h_\lostB + \zeta^d_\lostB)}{p^1_\sent - \zeta^1_\sent}\right)\right)M
\right)\ .
\label{eq:thmepsdef}
\end{align}
\item
Suppose $\cF=\cN^{\otimes \nu M_{\rm store}}$ for a storage rate~$\nu>0$, $\cN$ satisfying the strong converse property~\cite{rs:converse}
and
having capacity~$C_\cN$ bounded by
\begin{align}
C_\cN\cdot\nu< \min_{r^{(1)}} R\ .
\end{align} 
Let $\delta\in ]0,\frac{1}{2}-C_\cN\cdot \nu[$.
Then we obtain a min-entropy rate of
\begin{align}
\tilde{\lambda}(\delta) &=
\min_{\{r^{(n)}\}_n}
\frac{1}{m} 
\left[
\nu \cdot \gamma^\cN\left(\frac{R}{\nu}\right) M_{\rm store} - 
\sum_{n=2}^{\infty} M^{(n)}_{\rm left} \log\left(1 - p^{d,n}_{\errB}\right)\right]\ ,
\end{align}
for sufficiently large~$M$.
\end{enumerate}
\end{theorem}

\section{Proof of security: FROT from WSEE}
\label{app:OT}

We show that our augmented protocol implements fully randomized
oblivious transfer, as defined in~\cite{noisy:new}. The proofs of
correctness and security for honest Bob are analogous to the ones
given in~\cite{noisy:new}, using the fact that the properties of the
error-correcting code ensure that Bob obtains $S_C$ except with
probability $\eps$. Furthermore, note that a dishonest Alice cannot
gain any information about $C$ from a one-way error-correction
scheme. We therefore concentrate on proving security for an honest
Alice when Bob is dishonest. The proof proceeds as
in~\cite{noisy:new}, except for a small variation which we state
below.

\begin{lemma} [Security for Alice] Let $\ell := \left \lfloor
    \left(\left(\frac{\omega-1}{\omega}\right)\frac{\lambda}{8}-\frac{\lambda^2}{512 \omega^2\beta} - \frac{1.2
        \cdot h(p_{\rm err})}{8}\right)m - \frac{1}{2} \right
  \rfloor$. Then, Protocol WSEE-to-FROT satisfies security for Alice
  with an error of
\[41 \cdot 2^{-\frac{\lambda^2}{512 \omega^2\beta} m} + 2\varepsilon\;.\]
\end{lemma}

\begin{proof}
We know from the analysis in~\cite{noisy:new} that
\begin{align}\label{eq:prevAnalysis}
\hmin^{\varepsilon+4\delta}(\Pi({\bf Z})_{\Enc(W_{1-C}^t)}| S_{C}^\ell C
R_0 R_1 W_0^t W_1^t \Pi B''', \cA)_{\tilde \sigma}
\geq \left( \frac{\omega-1}{\omega} \right) \frac{\lambda m}{4} - \ell - 1\; ,
\end{align}
where $B'''$ is the system of dishonest Bob after the interactive
hashing protocol and $\cA$ is the event that the interactive hashing
protocol provides us with a set $W^t_{1-C}$ of high min-entropy. $\cA$
has probability $\Pr[\cA] \geq 1 - 32 \delta^2$, where $\delta =
2^{-\alpha \lambda^2/(512 \omega^2)}$.  Here, Bob has some additional information
given by the syndromes $\Syn(\Pi({\bf Z})_j)$ of the blocks $j \in
{\Enc(W_{0}^t)} \cup {\Enc(W_{1}^t)}$. Let us denote the total of this
error-correcting information by $\Syn := \{ \Syn(\Pi({\bf Z})_j) \}_{j
  \in \Enc(W_{0}^t) \cup \Enc(W_{1}^t)}$. Notice that even if the
encodings overlap in some blocks, only the syndromes of the $\alpha/4$
blocks in $\Enc(W_{1-C}^t)$ lower Bob's min-entropy on $\Pi({\bf
  Z})_{\Enc(W_{1-C}^t)}$. We can hence bound
\begin{align}
&\hmin^{\varepsilon+4\delta}(\Pi({\bf Z})_{\Enc(W_{1-C}^t)}| S_{C}^\ell C
R_0 R_1 W_0^t W_1^t \Pi \Syn B''', \cA)_{\tilde \sigma}\\
&\qquad\geq \hmin^{\varepsilon+4\delta}(\Pi({\bf Z})_{\Enc(W_{1-C}^t)}| S_{C}^\ell C
R_0 R_1 W_0^t W_1^t \Pi B''', \cA)_{\tilde \sigma} - 1.2 \cdot h(p_{\rm err}) \frac{m}{4}\\
&\qquad\geq \left(\left(\frac{\omega-1}{\omega}\right)\frac{\lambda}{4} -  \frac{1.2 \cdot h(p_{\rm err})}{4}\right) m - \ell - 1\; ,
\end{align}
where the first inequality follows from the chain rule, the
monotonicity of the smooth min-entropy~\cite{renato:diss}, and the fact
that error-correction information needs to be send for $\beta \cdot \alpha/4
= m/4$ bits.  Using privacy amplification~\cite{renato:diss}, we then
have that, conditioned on the event $\cA$,
\begin{align}
\frac{1}{2}
\|\tilde{\sigma}_{S_{1-C},S_{C} C R_0 R_1 W_0^t W_1^t \Pi \Syn B'''} - 
\tau_{\sbin^\ell} \otimes \tilde{\sigma}_{S_{C} C R_0 R_1 W_0^t W_1^t
  \Pi \Syn B'''}\|_1
\leq \delta +
2\varepsilon+8\delta\;,
\end{align}
since
\[
\left(\frac{\omega-1}{\omega}\right)\frac{\lambda m}{4} - \frac{1.2 \cdot h(p_{\rm err})m}{4}- 2\ell - 1\geq 2 \log 1/\delta = 2 \cdot
\frac{ \lambda^2 \alpha}{512 \omega^2}\;,\] which follows from
\[\ell \leq \left(\left(\frac{\omega-1}{\omega}\right)\frac{\lambda}{8}  - \frac{1.2 \cdot h(p_{\rm err})}{8}\right) m -  \frac{ \lambda^2 \alpha}{512 \omega^2} - \frac{1}{2}\;.\]
Let $B^* := (R_0 R_1 W_0^t W_1^t \Pi \Syn B''')$ be Bob's part
in the output state. Since
$\Pr[\cA] \geq 1- 32 \delta^2$, we get
\[
 \tilde \sigma_{S_{1-C} S_C B^* C} \approx_{32 \delta^2 +9\delta +
2\varepsilon} \tau_{\{0,1\}^\ell} \otimes \tilde \sigma_{ S_C B^* C}
\]
and
\[
 \tilde \sigma_{S_0 S_1 B^*} = \tilde \rho_{S_0 S_1 B^*}\;.
\]
Since $\delta^2 \leq \delta$, this implies the security condition for
Alice, with a total error of at most $41 \delta + 2\varepsilon$.
\end{proof}

\section{Derivation of parameters}

In this section, we show how to compute the parameters for both experimental setups. 

\subsection{Weak coherent source}

The case of phase-randomized weak coherent pulses is particularly easy to analyze,
since here we can assume that Bob always gains full knowledge of the encoded
bit from a multi-photon emission. That is, $p^{d,n}_\errB = 0$ for all $n>1$.  In
particular, this yields
\begin{align}
p^d_\lostB = p^0_\src = e^{-\mu},
\end{align}
and
\begin{align}
p^1_\src = e^{-\mu} \mu. 
\end{align}
The action of Bob's detection device can be described by two {\it
  positive-operator valued measures} (POVM), one for each of the two
polarization bases $\beta$ used in the BB$84$ protocol. Each POVM
contains four elements: $F_{\rm vac}^\beta, F_{\rm 0}^\beta, F_{\rm
  1}^\beta$, and $F_{\rm D}^\beta$. The outcome of the first operator,
$F_{\rm vac}^\beta$, corresponds to no click in the detectors, the
following two POVM operators, $F_{\rm 0}^\beta$ and $F_{\rm 1}^\beta$,
give precisely one detection click, and the last one, $F_{\rm
  D}^\beta$, gives rise to both detectors being triggered. If we
denote by $\ket{n,m}_\beta$ the state which has $n$ photons in one
mode and $m$ photons in the orthogonal polarization mode with respect
to the polarization basis $\beta$, the elements of the POVM for this
basis are given by
\begin{eqnarray}\label{det}
F_{vac}^\beta&=&\sum_{n,m=0}^{\infty}\ \bar{\eta}^{n+m}\
\ket{n,m}_\beta\bra{n,m},
\\F_{0}^\beta&=&\sum_{n,m=0}^{\infty}\ (1-\bar{\eta}^n)\bar{\eta}^{m}\
\ket{n,m}_\beta\bra{n,m}, \nonumber\\
F_{1}^\beta&=&\sum_{n,m=0}^{\infty}\
(1-\bar{\eta}^m)\bar{\eta}^{n}\
\ket{n,m}_\beta\bra{n,m},\nonumber\\
F_D^\beta&=&\sum_{n,m=0}^{\infty}\
(1-\bar{\eta}^n)(1-\bar{\eta}^m)\ \ket{n,m}_\beta\bra{n,m},
\nonumber
\end{eqnarray}
where $\eta$ is the detection efficiency 
of a detector
as introduced in
Section~\ref{sec_wcp_pract} and $\bar{\eta}=(1-\eta)$.  Furthermore,
we take into account that the detectors show noise in the form of dark
counts which are, to a good approximation, independent of the incoming
signals. As in Section~\ref{sec_wcp_pract}, the dark count probability
of each detector is denoted by $p_{\rm dark}$.

First of all, since Alice does not verify how many photons have actually been emitted we have
\begin{align}
p^n_\sent = p^n_\src\ .
\end{align}
To determine the other parameters, we start by
computing the probability that an honest Bob
does not observe a click due to a signal being sent which can be expressed as
\begin{align}
p^h_{\slostB} = \Tr(F^\beta_{\rm vac}\rho_k) = e^{-\mu} \sum_{n=0}^{\infty} \frac{\mu^n}{n!} (1-\eta)^n\ ,
\end{align}
with $\rho_k$ given by \eqref{mv}.
Conversely, the probability that Bob does see a click due to a signal being sent is 
\begin{align}
p^h_{\sclickB} = 1 - p^h_{\slostB}.
\end{align}
To calculate the total probability of Bob observing a click in his detection apparatus, we have to take dark counts
into account. 
We now write the probability of Bob observing no-click due to a dark count as
\begin{align}
p_{\dlostB} = (1- p_{\dark})^2,
\end{align}
and the probability that at least one of his two detectors clicks becomes
\begin{align}
p_{\dclickB} = p_{\dark} (2 - p_{\dark}).
\end{align}
The total probability that honest Bob observes a click is thus
\begin{align}
p^h_{\clickB} = p^h_{\sclickB} p_{\dlostB} + p^h_{\slostB} p_{\dclickB} + p^h_{\sclickB} p_{\dclickB} = 
p^h_{\sclickB} + p^h_{\slostB} p_{\dclickB}.
\end{align}
Note that 
\begin{align}
p^h_{\lostB} = 1 - p^h_{\clickB}\ .
\end{align}

To finish our analysis, it remains to evaluate the error probability
for honest Bob, which determines how much error-correcting information
Alice will send him. First of all, an error may occur from the signal
itself, for example due to misalignment in the channel.  We have
\begin{align}
p^h_\serr = \e_{\rm det} \cdot p^h_{\sclickB}\ .
\end{align}
The second source of errors are dark counts. If the signal has been lost, 
the probability of making an error due to a dark count is given by the 
probability that Bob experiences a click in the wrong detector, or both his detectors click. Hence, we have
\begin{align}
p_\derr = p_\dark (1-p_\dark) + p_\dark^2/2,
\end{align}
where the second term stems from letting Bob flip a coin to determine
the outcome bit when both of his detectors click.  We can also have a
combination of errors from the signal and the dark counts. Considering
all different possibilities we obtain
\begin{align}
p_\dserr = p^h_{\sclickB} \left((1 - \edet)\frac{p_\dark}{2} + \edet p_\dark \left(\frac{3}{2} - p_\dark\right)\right).
\end{align}
Putting everything together we have
\begin{equation}\label{eq:totalErr}
p^h_{\errB} = p^h_\serr p_{\dlostB} + p^h_{\slostB} p_\derr + p_\dserr.
\end{equation}

\subsection{Parametric down conversion source}

In this section, we show how to compute all relevant parameters for a
PDC source.  Recall that at each time slot, the source itself emits
an entangled state given by~(\ref{marcos1}). The state $\ket{\Phi_n}_{AB}$ which 
appears in~(\ref{marcos2}) can be written as 
\begin{align}
\ket{\Phi_n}_{AB} = \sum_{m=0}^n \frac{(-1)^m}{\sqrt{n+1}} \frac{(a_1^\dagger)^{n-m}}{\sqrt{(n-m)!}} \frac{(a_2^\dagger)^m}{\sqrt{m!}}\ket{0,0}_A \ket{m,n-m}_B.
\end{align}
\begin{figure}[h]
\begin{center}
\includegraphics[width=3cm]{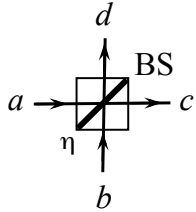}
\caption{$a$ and $b$ denote the input modes to a beam splitter (BS) of transmittance $\eta$, while $c$ and $d$ are 
the output modes. }
\label{bs_m}
\end{center}
\end{figure}

We shall consider that both detectors on Alice's side are equal. In this situation, it is possible to attribute their losses 
to a single-loss beam splitter of transmittance $\eta$ as illustrated in Figure~\ref{bs_m}. 
The creation operators $a_1^\dagger$ and $a_2^\dagger$ can be expressed as
\begin{eqnarray*}
a_1^\dagger &=& \sqrt{\eta} c_1^\dagger + \sqrt{1-\eta} d_1^\dagger,\\
a_2^\dagger &=& \sqrt{\eta} c_2^\dagger + \sqrt{1-\eta} d_2^\dagger,
\end{eqnarray*}
for the two orthogonal polarization modes. Tracing out the modes $d_1$ and $d_2$ 
we obtain that the state shared by Alice and
Bob, after accounting for Alice's losses, is given by
\begin{eqnarray*}
\rho_{AB} 
&=& \sum_{n,n'} \sqrt{p^n_\src p^{n'}_\src} \sum_{m=0}^n \sum_{m'=0}^{n'} \sum_{j=0}^{\min(n-m,n'-m')} \sum_{\ell=0}^{\min(m,m')}
\sqrt{\frac{(n-m)! m!}{(n-m-j)! j! (m-\ell)! \ell!}}\\
&&
\sqrt{\frac{(n'-m')! m'!}{(n'-m'-j)! j! (m'-\ell)! \ell!}}
\frac{(-1)^{m+m'}}{\sqrt{n+1}\sqrt{n'+1}} \sqrt{\eta}^{n+n' - 2j - 2\ell} \sqrt{1-\eta}^{2(j+\ell)} \\
&&\outp{n-m-j,m-\ell}{n'-m'-j,m'-\ell}_A \otimes
\outp{m,n-m}{m',n'-m'}_B.
\end{eqnarray*}
Even though we again have two bases of course, we will only consider one of the two, the other one merely differs in a prior transform by Alice
and does not change the resulting probabilities. 
For perfect threshold detectors, the probability that Alice sees a click in her first detector (concluding an encoding of '0') is given by
\begin{align}
p^0_{\sclickA} = \Tr((C_1^A \otimes \id^B)\rho_{AB}) = \sum_{n=1}^{\infty} \frac{p^n_\src}{n+1},
\sum_{m=0}^{n-1} [(1-\eta)^m - (1-\eta)^n]
\end{align}
where 
\begin{align}
C_1^A = \sum_{n=1}^{\infty} \proj{n}_{c_1} \otimes \proj{0}_{c_2}.
\end{align}
The probability that she observes a click in the second detector is similarly determined by $p^1_{\sclickA}=\Tr((C_2^A \otimes \id^B)\rho_{AB})$
with
\begin{align}
C_2^A = \proj{0}_{c_1} \otimes \sum_{n=1}^{\infty} \proj{n}_{c_2}.
\end{align}
If Alice sees no click in a given round, or both her detectors click,
she simply discards this round all together and it no longer
contributes to the protocol.  We have that $p^0_{\sclickA} = p^1_{\sclickA}$. 

As discussed previously, we consider that the noise in the form of dark counts shown by the detectors is, to 
a good approximation, independent of the incoming signals. Then, to include this effect, 
we have to consider the probability of observing a click due to a dark count alone. This is given by the probability 
that we detect no photons
\begin{align}
p_{\rm vac} = \Tr((\proj{0,0}_{c_1,c_2} \otimes \id^B)\rho_{AB}),
\end{align}
but the detector clicks because of a dark count.
We can obtain the probability that Alice observes only one click due to a 
signal or a dark count, by considering operators of the form
\begin{eqnarray*}
\hat{C}_1^A &=& (1-p_\dark) C_1^A + (1-p_\dark)p_\dark \proj{0,0}_{c1,c2},\\
\hat{C}_2^A &=& (1-p_\dark) C_2^A + (1-p_\dark)p_\dark \proj{0,0}_{c1,c2},
\end{eqnarray*}
which gives us
\begin{align}
p^0_{\clickA} = p^1_{\clickA} = (1-p_\dark) p^0_{\sclickA} + (1-p_\dark)p_\dark \sum_{n=0}^\infty p^n_\src (1-\eta)^n.
\end{align}
Combining everything, and tracing out Alice's register we obtain that Bob's unnormalized states are given by
\begin{eqnarray*}
\tilde{\rho}^{0}_{B} &=& (1-p_\dark) \tilde{\rho}^{0,\sig}_{B} + (1 - p_\dark)p_\dark \tilde{\rho}^{\vac}_{B},\\
\tilde{\rho}^{1}_{B} &=& (1-p_\dark) \tilde{\rho}^{1,\sig}_{B} + (1 - p_\dark)p_\dark \tilde{\rho}^{\vac}_{B},
\end{eqnarray*}
with
\begin{eqnarray*}
\tilde{\rho}^{0,\sig}_{B} &=& \sum_{n=1}^{\infty} \frac{p^n_\src}{n+1} \sum_{m=0}^{n-1} \left[(1-\eta)^m - (1-\eta)^n\right]
\outp{m,n-m}{m,n-m}_{B},\\
\tilde{\rho}^{1,\sig}_{B} &=& \sum_{n=1}^{\infty} \frac{p^n_\src}{n+1} \sum_{m=0}^{n-1} \left[(1-\eta)^m - (1-\eta)^n\right]
\outp{n-m,m}{n-m,m}_{B},\\
\tilde{\rho}_{B}^{vac}&=&\sum_{n=0}^{\infty} \frac{p_{src}^n
(1-\eta)^n}{n+1}  \sum_{m=0}^{n}\ket{m,n-m}\bra{m,n-m}_B,
\end{eqnarray*}
In the following, we use $\rho = \tilde{\rho}/\Tr(\tilde{\rho})$ to refer to the normalized versions of these states.
Note that these normalization factors are the same for an encoding of a '0' or a '1' and are given by 
$c=p^0_{\clickA}$.

We can now write the probability that the source emits $n$ photons given that Alice obtained one single click in 
her measurement apparatus as 
\begin{align}\label{eq:pdcSent}
p^n_{\sent} := \frac{1}{c} p^n_{\src} 
(1-\pdark) \left(\pdark (1-\eta)^n 
+\frac{1}{n+1}
\sum_{m=0}^n \left((1-\eta)^m - (1-\eta)^n\right)\right)\ .
\end{align}

We are now ready to compute the probabilities relevant to the security analysis.
First of all, we need to know the probability that honest Bob observes a click for the 
pulses where Alice has obtained one single click, 
\begin{align}
p^h_{\clickB} = p^h_{\sclickB} p_{\dlostB} + p^h_{\slostB} p_{\dclickB} + p^h_{\sclickB} p_{\dclickB}.
\end{align}
The probability that honest Bob does not observe a click at all, due to the signal is given by
\begin{eqnarray*}
p^{h}_{\slostB} &=& \Tr(F^{\beta}_{\vac}\rho^0_B)\\
&=& \frac{1}{c} \left[ p_\dark (1-p_\dark) \sum_{n=0}^{\infty} p^n_\src (1-\eta)^{2n} + (1-p_\dark) \sum_{n=0}^{\infty}
\frac{p^n_\src}{n+1}\sum_{m=0}^n \left[(1-\eta)^m - (1-\eta)^n\right] (1-\eta)^n\right],
\end{eqnarray*}
and
\begin{equation}\label{mm}
p^h_{\sclickB} = 1 - p^h_{\slostB},
\end{equation}
where the probabilities 
$p_{\dlostB} $ and $p_{\dclickB}$
are defined in the same way as
in the previous section.  We also need to determine the probability
of an error for honest Bob. This is calculated analogous to the case
of a weak coherent source, where we consider the probabilities of an
error due to the signal itself, dark counts, and both combined.  In
our setting an honest Bob has two detectors to decide what bit Alice
has encoded. If both detectors click, we shall consider again that honest Bob flips a coin to
determine the outcome. It is enough to analyze the case of a '0'
encoding; the  '1' encoding provides the same result. 
The probability that Bob makes an error due to the signal
is given by
\begin{align}\label{eq:pdcSignalErr}
p^h_{\serrB} = \frac{1}{c} \Tr(F \tilde{\rho}^0_B),
\end{align}
where
\begin{eqnarray*}
F &=& \tilde{F}_0^\beta + \frac{1}{2} F^{\beta}_D,\\
\tilde{F}_0^\beta &=& (1 - \edet)F_0^\beta + \edet F_1^\beta,
\end{eqnarray*}
and $F_0^\beta$, $F_1^\beta$, and $F_D^\beta$ are given by~(\ref{det}).
Note
that
\begin{align}\label{eq:pdcSignalClick}
p^h_{\sclickB} = p^h_{\serrB} + p^h_{\snoerrB}.
\end{align}
Then, using that
\begin{equation}\label{mm2}
p_{B,DS,\rm err}=p_{B,S,\rm err}^h \pdark \Big(\frac{3}{2}-\pdark\Big)+p_{B,S,no\ err}^h \frac{\pdark}{2},\, 
\end{equation}
we can now compute the combined error of Bob as in Eq.~(\ref{eq:totalErr}).

In the case of PDC source we also need to compute Bob's success probability
of decoding a bit from a multi-photon emission, if he is given the
basis information for free.  First of all, note that since
$\rho^0_{B}$ and $\rho^1_{B}$ are Fock diagonal states, without loss
of generality we can always assume that dishonest Bob first measures the photon number of each pulse sent by Alice,
and afterwards he performs his attack.  For $n \geq 1$, we have
\begin{eqnarray*}
\tilde{\rho}^{0,n,\sig}_B &=&
\frac{p_{src}^n}{n+1}
\sum_{m=0}^{n-1} [(1-\eta)^m - (1-\eta)^n] \outp{m,n-m}{m,n-m}_B,\\
\tilde{\rho}^{1,n,\sig}_B &=&
\frac{p_{src}^n}{n+1}
\sum_{m=0}^{n-1} [(1-\eta)^m - (1-\eta)^n] \outp{n-m,m}{n-m,m}_B,\\
\tilde{\rho}^{\vac,n}_B &=& p_{\src}^n \frac{(1-\eta)^n}{n+1} \sum_{m=0}^{n} \outp{m,n-m}{m,n-m}_B.
\end{eqnarray*}
The unnormalized states of Bob containing $n$ photons and corresponding to an encoding of a '0'
or '1' respectively can then be written as
\begin{eqnarray*}
\tilde{\rho}^{0,n}_B &=& (1 - \pdark) \tilde{\rho}^{0,n,\sig}_B + (1-\pdark)\pdark \tilde{\rho}^{\vac,n}_B,\\
\tilde{\rho}^{1,n}_B &=& (1 - \pdark) \tilde{\rho}^{1,n,\sig}_B + (1-\pdark)\pdark \tilde{\rho}^{\vac,n}_B.
\end{eqnarray*}
The normalization factor for both states is
\begin{eqnarray*}
c_n &\assign& \Tr(\tilde{\rho}^{0,n}_B) = (1 -\pdark) \Tr(\tilde{\rho}^{0,n,\sig}_B) + (1-\pdark) \pdark \Tr(\tilde{\rho}^{\vac,n}_B) \\
&=& 
(1-\pdark)\frac{p_{src}^n}{n+1} \sum_{m=0}^n [(1-\eta)^m - (1-\eta)^n] + (1-\pdark) \pdark p_{src}^n(1-\eta)^n\ .
\end{eqnarray*}

\begin{claim}\label{lem:comparison}
The probability that Bob makes an error in decoding if Alice sent an $n$-photon signal and he is given the basis information for 
free is given by
\begin{equation}\label{mm3}
p^{d,n}_{\errB} = \frac{1}{2} - \frac{1}{4}\left(\frac{1-\pdark}{c_n}
\frac{p_{src}^n}{n+1}  \sum_{m=0}^n |(1-\eta)^m - (1-\eta)^{n-m}|\right).
\end{equation}
\end{claim}
\begin{proof}
  This is an immediate consequence of Helstrom's
  theorem~\cite{helstrom:detection} using the fact that an encoding of
  '0' and '1' are a priori equally probable for Bob. Furthermore, note that 
  $\rho^{0,n}_B$ 
and  $\rho^{1,n}_B$
are both Fock diagonal, and hence their trace distance is simply given by the classical statistic distance on the r.h.s. of
\begin{align}
\frac{1}{2} ||\rho^{0,n}_B - \rho^{1,n}_B||_1 = \frac{1-\pdark}{2 c_n} 
\frac{p_{src}^n}{n+1} 
\sum_{m=0}^n |(1-\eta)^m - (1-\eta)^{n-m}|.
\end{align}
\end{proof}

\end{document}